\documentclass[superscriptaddress,prx,twocolumn]{revtex4-1}
\usepackage{graphicx}
\usepackage{amsmath,amssymb,physics,dsfont}
\usepackage{bm}
\usepackage{mathrsfs}
\usepackage{algpseudocode}
\usepackage{algorithm}
\usepackage{color}
\usepackage{hyperref}
\usepackage{microtype}
\usepackage{braket}
\usepackage{defs-private}

\hypersetup{
    unicode=false,          
    pdftoolbar=true,        
    pdfmenubar=true,        
    pdffitwindow=false,     
    pdfstartview={FitH},    
    pdftitle={},    
    pdfauthor={},     
    pdfsubject={},   
    pdfcreator={},   
    pdfproducer={}, 
    pdfkeywords={} {} {}, 
    pdfnewwindow=true,      
    colorlinks=true,       
    linkcolor=magenta, 
    citecolor=blue,        
    filecolor=magenta,      
    urlcolor=blue           
}

\begin{document}

\title{Comparative study of adaptive variational quantum eigensolvers for multi-orbital impurity models}

\author{Anirban Mukherjee}
\affiliation{Ames National Laboratory, U.S. Department of Energy, Ames, Iowa 50011, USA}

\author{Noah F. Berthusen}
\affiliation{Ames National Laboratory, U.S. Department of Energy, Ames, Iowa 50011, USA}
\affiliation{Department of Electrical and Computer Engineering, Iowa State University, Ames, Iowa 50011, USA}
\affiliation{Department of Computer Science, University of Maryland, College Park, MD, 20742, USA}

\author{Jo\~{a}o C. Getelina}
\affiliation{Ames National Laboratory, U.S. Department of Energy, Ames, Iowa 50011, USA}

\author{Peter P. Orth}
\email{porth@iastate.edu}
\affiliation{Ames National Laboratory, U.S. Department of Energy, Ames, Iowa 50011, USA}
\affiliation{Department of Physics and Astronomy, Iowa State University, Ames, Iowa 50011, USA}

\author{Yong-Xin Yao}
\email{ykent@iastate.edu}
\affiliation{Ames National Laboratory, U.S. Department of Energy, Ames, Iowa 50011, USA}
\affiliation{Department of Physics and Astronomy, Iowa State University, Ames, Iowa 50011, USA}

\begin{abstract}
~\\
\begin{center}
\textbf{Abstract}
\end{center}
    Hybrid quantum-classical embedding methods for correlated materials simulations provide a path towards potential quantum advantage. However, the required quantum resources arising from the multi-band nature of $d$ and $f$ electron materials remain largely unexplored. Here we compare the performance of different variational quantum eigensolvers in ground state preparation for interacting multi-orbital embedding impurity models, which is the computationally most demanding step in quantum embedding theories. Focusing on adaptive algorithms and models with 8 spin-orbitals, we show that state preparation with fidelities better than $99.9\%$ can be achieved using about $2^{14}$ shots per measurement circuit. When including gate noise, we observe that parameter optimizations can still be performed if the two-qubit gate error lies below $10^{-3}$, which is slightly smaller than current hardware levels. Finally, we measure the ground state energy on IBM and Quantinuum hardware using a converged adaptive ansatz and obtain a relative error of 0.7\%.
    
\end{abstract}

\maketitle

\section{Introduction}
Eigenstate preparation for Hamiltonian systems is one promising application of noisy intermediate-scale quantum (NISQ) computers to achieve practical quantum advantage~\cite{asp_ipea, peruzzoVariationalEigenvalueSolver2014, hardware_efficient_vqe, vqe_pea_h2, nisq, rmp_qcc, cerezo2021variational}. One of the representative hybrid quantum-classical algorithms to achieve this task is the variational quantum eigensolver (VQE). It attempts to find the ground state of a given Hamiltonian $H$ within a variational manifold of states that are generated by parametrized quantum circuits $U(\bm{\theta})$ acting on a reference state $\ket{\Psi_0}$. The parameters $\bm{\theta}$ are obtained by classically minimizing the energy cost function $E(\bm{\theta}) = \braket{\Psi_0| U^\dag(\bm{\theta})HU(\bm{\theta})|\Psi_0}$ that is measured on quantum hardware~\cite{peruzzoVariationalEigenvalueSolver2014, hardware_efficient_vqe, vqe_pea_h2, vqe_theory}. The quality of a VQE calculation is tied to the ability of the variational ansatz to represent the ground state with high fidelity. In quantum computational chemistry, the unitary coupled cluster ansatz truncated at single and double excitations (UCCSD) has been extensively studied, owing to the success of the classical coupled cluster algorithm~\cite{hoffmann1988unitary, bartlett1989alternative, rmp_cc}. It was found that the application of UCCSD ansatz is limited by the rapid circuit growth with system size and the deteriorating accuracy in the presence of static electron correlations~\cite{vqe_theory, alan_ucc2018, grimsleyAdaptiveVariationalAlgorithm2019}. Therefore, alternative variants have been developed, including hardware-efficient ans\"atze, that improve the trainability and expressivity of the wave function ansatz~\cite{hardware_efficient_vqe, qcc_scott2018, kUpUCCGSD, grimsleyAdaptiveVariationalAlgorithm2019, MayhallQubitAVQE,FengVQE, AVQITE, fedorovVQEMethodShort2022, tilly2021variational}. 

Indeed, it was found that compact and numerically exact variational ground state ans\"atze can be \emph{adaptively} constructed for specific problems using approaches like the adaptive derivative-assembled pseudo-trotter (ADAPT) ansatz~\cite{grimsleyAdaptiveVariationalAlgorithm2019,MayhallQubitAVQE}. The adaptive ansatz is typically obtained by successively appending parametrized unitaries to a variational circuit with generators chosen from a predefined operator pool. In practice, the ADAPT-VQE algorithm works well with an operator pool composed of fermionic excitation operators in the UCCSD ansatz. The extended qubit-ADAPT VQE approach~\cite{MayhallQubitAVQE} utilizes an operator pool composed of Pauli strings in the qubit representation of fermionic excitation operators in the UCCSD ansatz, which is shown to be capable of generating significantly more compact ans\"atze than the original ADAPT-VQE method at the price of introducing more variational parameters. As the circuit complexity (i.e., the number of two-qubit operations in the circuit) is a determining factor for practical calculations on NISQ devices, qubit-ADAPT is preferable and chosen for the comparative study in this work. Regarding the scalability of the qubit-ADAPT method towards larger system sizes, we note that reference~\cite{AVQITE} reports a favorable linear system-size scaling for the adaptive ansatz complexity of nonintegrable mixed-field Ising model using the adaptive variational quantum imaginary time evolution method (AVQITE). AVQITE is known to generate variational circuits of comparable complexity as qubit-ADAPT VQE. As a first step to investigate the scalability in fermionic models, we here study qubit-ADAPT VQE for fermionic models with two and three spinful orbitals. 

An alternative approach of constructing efficient wavefunction ans\"atze for problems in condensed matter physics is to exploit the sparsity of the Hamiltonian. Interacting electron systems are often simulated with reduced degrees of freedom, represented, for example, by a single-band Hubbard model. This simplified model features a sparse Hamiltonian including nearest-neighbor hopping and onsite Coulomb interactions only. Motivated by the simplicity of the Trotterized circuits for dynamics simulations due to Hamiltonian sparsity, the Hamiltonian variational ansatz (HVA) has been proposed by promoting the time in Trotter circuits to independent variational parameters~\cite{wecker2015_trotterizedsp}. The HVA ansatz has attracted much attention and turns out to be very successful in reaching a compact state representation for sparse Hamiltonian system including local spin models~\cite{wecker2015_trotterizedsp, ho2019efficient, wiersema2020exploring}. Here, we propose to combine the flexibility of an adaptive approach with the efficiency of the HVA by designing a ``Hamiltonian commutator'' (HC) operator pool that contains pairwise commutators of operators that appear in the Hamiltonian. 

To obtain a realistic description of correlated quantum materials, which typically contain partially filled $d$-orbitals such as transition metal compounds, or $f$-orbitals such as rare-earth and actinide systems, it is important to go beyond the single-orbital description of a simple Hubbard model~\cite{kent2018toward}. Intriguing physics arises from the local Hund's coupling of electrons in different atomic orbitals. Examples are bad metallic behaviour with suppressed quasiparticle coherence and orbital-selective Mott transitions or superconducting pairing, which naturally require a multi-orbital description~\cite{Yin2011KineticFA, georgesStrongCorrelationsHund2013, deMedici2014SelectiveMP, Sprau2017DiscoveryOO}. A multi-orbital model including additional inter-orbital hoppings and Hund's couplings will necessarily make the Hamiltonian less sparse and consequently the HVA ansatz more complicated. Nevertheless, the complexity of material simulations can be greatly reduced by quantum embedding methods which maps the infinite system to coupled subsystems, typically a noninteracting effective medium and some many-body interacting impurity models~\cite{kent2018toward, dmft_georges96, dmft_kotliar06, sun2016quantum, knizia2012dmet, ga_uo2, ga_dmet, gqce, sakurai2021hybrid, Vorwerk2022QuantumET}. These quantum embedding approaches have proven to be very effective to simulate correlated electron systems, including energies, electronic structure, magnetism, superconductivity, and spectral properties of multiple competing phases. The computational load in these approaches is shifted from the solution of a full lattice system to that of an interacting multi-orbital impurity model. Classical algorithms for solving the impurity problem, however, are not scalable, which can be more tractable with quantum computers~\cite{gqce, hybrd_dmft}.

In this paper, we compare the VQE circuit complexity for ground state preparation of multi-orbital many-body impurity models with a fixed HVA versus a qubit-ADAPT ansatz with different operator pools. An HC operator pool compatible with HVA is proposed to allow a fair comparison between qubit-ADAPT and fixed ansatz HVA calculations. For comparison, we also include results from UCCSD and qubit-ADAPT calculations with a simplified UCCSD pool. To connect with quantum embedding methods for realistic materials simulations, we use the Gutzwiller embedding approach~\cite{ga_uo2, ga_bunemann1998, ga_fabrizio07, deng2008lda+, ga_ce, ga_smb6, ga_pu} to generate the impurity models that we employ for our benchmark~\cite{gqce, pygqce}. The quantum calculation we perform is general and could also be applied to other embedding methods. Numerical results from noiseless statevector simulator and quantum assembly language (QASM)-based simulator with quantum sampling noise are presented. Important techniques for efficient circuit simulations of qubit-ADAPT VQE are discussed, including ways to simplify generators and to reduce the operator pool size. We further investigate the impact of realistic gate noise by performing qubit-ADAPT VQE simulations with a realistic noise model including amplitude and dephasing channels. Finally, we measure the energy cost function of the converged VQE ansatz for the $e_g$ model composed of $8$ spin-orbitals on the IBM quantum processing unit (QPU) \texttt{ibmq\_casablanca} and on Quantinuum hardware.


\begin{figure}
	\centering
	\includegraphics[width=\linewidth]{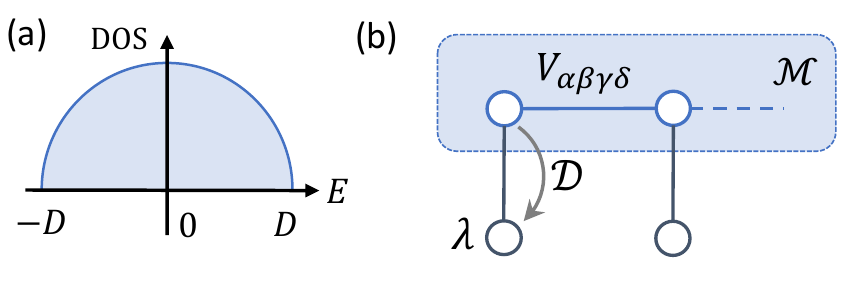}
	\caption{
	\textbf{Model setup.} (a) The noninteracting density of states (DOS) of the degenerate multi-band Hubbard-Hund lattice model on the Bethe lattice has a semicircular shape. (b) ($\M, \M$) site impurity model with $\M$-fold degenerate correlated orbitals coupled with $\M$ bath orbitals. The interactions among the physical orbitals are specified by the Coulomb matrix $V$. Due to symmetry, each physical orbital (positioned at zero energy level) is coupled with a single bath orbital at energy level $\lambda$ with a coupling parameter $\D$. The models with $\M=2$ and $3$ correspond to that of $e_g$ and $t_{2g}$ orbitals in cubic crystal symmetry, respectively.
	}
	\label{fig: model}
\end{figure}

\section{Results and discussion}
\subsection{Quantum embedding model} \label{sec: model}
Here we focus on a specific quantum embedding method: the well-established Gutzwiller variational embedding approach for correlated material simulations~\cite{ga_uo2, ga_bunemann1998, ga_fabrizio07, deng2008lda+, ga_ce, ga_smb6, ga_pu}, which is known to be equivalent to rotationally invariant slave-boson theory at the saddle point approximation~\cite{sb_kotliar86, bunemann_gasb}. Recently, our group has developed a hybrid Gutzwiller quantum-classical embedding approach (GQCE)~\cite{gqce}. GQCE maps the ground state solution of a correlated electron lattice system to a coupled eigenvalue problem of a noninteracting quasiparticle Hamiltonian and one or multiple finite-size interacting embedding Hamiltonians~\cite{ga_pu}. Within GQCE one employs a quantum computer to find the ground state energy and single-particle density matrix of the interacting embedding Hamiltonian, for example, using VQE. 

The embedding Hamiltonian describes an impurity model consisting of a physical many-body $N_\mathcal{S}$-orbital subsystem ($\h_\mathcal{S}$) coupled with a $N_\mathcal{B}$-orbital quadratic bath ($\h_\mathcal{B}$):
\be
\h = \h_\mathcal{S} + \h_\mathcal{B} + \h_\mathcal{SB}, \label{eq: h}
\ee
with 
\bea
&&\h_\mathcal{S} = \sum_{\alpha\beta}\sum_{\s} \epsilon_{\alpha\beta}\cc_{\alpha\s}\ca_{\beta\s} \notag \\
&& \qquad+ \frac{1}{2}\sum_{\alpha\beta\gamma\delta}\sum_{\s\s'}V_{\alpha\beta\gamma\delta} \cc_{\alpha\s} \cc_{\gamma\s'} \ca_{\delta\s'} \ca_{\beta\s}, \\
&&\h_\mathcal{B} = -\sum_{a b}\sum_{\s}\lambda_{a b} \fc_{a\s}\fa_{b\s} , \\
&&\h_\mathcal{SB} = \sum_{a \alpha}\sum_{\s}\left( \D_{a \alpha} \cc_{ \alpha\s} \fa_{a\s} + h.c.\right).
\eea
Here $\alpha, \beta, \gamma, \delta$ are composite indices for sites and spatial orbitals in the physical subsystem. Likewise, the bath sites and orbitals are labelled by $a, b$, and $\s$ is the spin index. The fermionic ladder operators $\ca$ and $\fa$ are used to distinguish the physical and bath orbital sites. 

The one-body component and two-body Coulomb interaction in the physical subsystem are specified by matrix $\epsilon$ and tensor $V$. The quadratic bath and its coupling to the subsystem are defined by matrix $\lambda$ and $\D$, respectively. Compared with typical quantum chemistry calculations, the embedding Hamiltonian is much sparser since the two-body interaction only exists between electrons in the physical subsystem. 

For clarification, we name the above defined embedding Hamiltonian system as ($N_\mathcal{S}, N_\mathcal{B}$) impurity model, where ($N_\mathcal{S}, N_\mathcal{B}$) are the number of spatial orbitals in the system and bath models. Within GQCE, the ground state solution of the embedding Hamiltonian at half electron filling is needed, which is achieved by a chemical potential absorbed in the one-body Hamiltonian coefficient matrices $\epsilon$ and $\lambda$ in Eq.~\eqref{eq: h}. 

In the numerical simulations presented here, we choose a Gutzwiller embedding Hamiltonian for the degenerate $\M$-band Hubbard model. The noninteracting density of states of the lattice model adopts a semi-circular form $\rho(\omega)=\frac{2\M}{\pi D}\sqrt{1-(\omega/D)^2}$ as shown in Fig.~\ref{fig: model}(a), which corresponds to the Bethe lattice in infinite dimensions. In the following, we set the half band width $D=1$ as the energy unit. In physical systems $D$ is of the order of a few eV. The Coulomb matrix $V$ takes the Kanamori form specified by Hubbard $U$ and Hund's $J$ parameters: $V_{\alpha\alpha\alpha\alpha} = U$, $V_{\alpha\alpha\beta\beta} = U - 2J$, and $V_{\alpha\beta\alpha\beta} = V_{\alpha\beta\beta\alpha} = J$ for $\alpha \neq \beta$. Here we have assumed spin and orbital rotational invariance (within the $e_g$ or $t_{2g}$ manifold) for simplicity and to limit the interaction parameter space. 

The embedding Hamiltonian, as illustrated in Fig.~\ref{fig: model}(b), is represented with $2\M$ spatial orbitals: $\M$ degenerate physical orbital plus $\M$ degenerate bath orbitals. The symmetry of the model reduces matrices $\epsilon$, $\lambda$ and $\D$ to single parameters proportional to identity. 

In the following, we set the electron filling for the lattice model to $\M+1$, which is one unit larger than half-filling, and fix the ratio of the Hund's to Hubbard interaction to $J/U=0.3$ and $U=7$. These parameters put the model deep in the correlation-induced bad metallic state, with physical properties distinct from doped Mott insulators~\cite{Yin2011KineticFA}. It represents a wide class of strongly correlated materials, such as iron pnictides and chalcogenides, where the Hund's coupling significantly reduces the low energy quasiparticle coherence scale~\cite{georgesStrongCorrelationsHund2013, deMedici2011JanusfacedIO, Lanata-Hunds_metals-PRB-2013}. The Hund's metal physics is far beyond a static mean-field description, and requires treating the localized and itinerant characters of electrons on equal footing, which can be realized in the quantum embedding approach adopted here. 

In calculations below, we consider $\M=2$ and $\M=3$, which correspond to $e_g$ and $t_{2g}$ orbitals in cubic crystal symmetry, respectively. The associated $(N_\mathcal{S}, N_\mathcal{B}) = (2, 2)$ and $(3, 3)$ impurity models have in total 8 and 12 spin-orbitals. The two models host nontrivial many-body ground states, and represent important checkpoints along the path to achieve practical quantum advantage in correlated materials simulations through hybrid quantum-classical embedding framework. In quantum simulations reported below, parity encoding which exploits the symmetry in total number of electrons and spin $z$-component is used to transform the fermionic Hamiltonian to qubit representation.

\subsection{Variational quantum eigensolvers} \label{sec: method}
GQCE leverages quantum computing technologies to solve for the ground state of the embedding Hamiltonian, specifically the energy and one-particle density matrix. Note that the ground state is always prepared at half-filling for the embedding system, which is determined by the Gutzwiller embedding algorithm and is independent of the actual electron filling of the physical lattice model~\cite{ga_pu, ga_uo2}. For this purpose, we benchmark multiple versions of VQE with fixed or adaptively generated ansatz to prepare the ground state of the above embedding Hamiltonian. We consider VQE calculations with fixed UCCSD ansatz and the associated qubit-ADAPT VQE using a simplified UCCSD operator pool. The calculations are naturally performed in the molecular orbital (MO) basis representation, where the reference Hartree-Fock (HF) state becomes a simple tensor product state and fermionic excitation operators can be naturally defined. However, using a MO representation comes at the cost of reducing the sparsity of the embedding Hamiltonian compared to the atomic orbital (AO) basis representation. To take advantage of the Hamiltonian sparsity in AO representation, we consider a generalized form of the HVA, and the associated qubit-ADAPT VQE with a modified HC operator pool. 

\begin{figure}[t]
	\centering
	\includegraphics[width=0.9\linewidth]{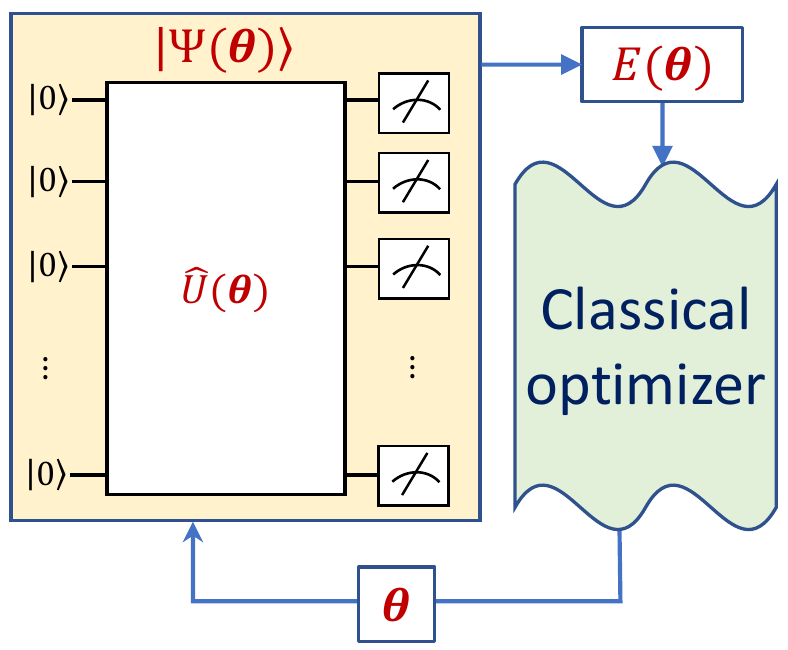}
	\caption{
	\textbf{Schematic illustration of the variational quantum eigensolver algorithm.} Given an initial guess for the parameter vector $\bth$, the many-body state is prepared using parametrized circuit $\hat{U}(\bth)$ on the quantum computer. A set of measurements are performed in the computational basis to estimate the cost function $E(\bth)$, possibly including classical postprocessing for error mitigation. This value is subsequently passed to a classical optimizer. The parameters $\bth$ are then updated by the optimizer, which triggers a new iteration of state preparation and energy measurement. The cycle continues until $E(\bth)$ converges.
	}
	\label{fig: vqe}
\end{figure}

\textbf{VQE algorithm.}
For an $N_q$-qubit system with Hamiltonian $\h$, VQE amounts to minimizing the cost function $E(\bth) = \Av{\Psi[\bth]}{\h}$ with respect to the variational parameters $\bth$, as schematically illustrated in Fig.~\ref{fig: vqe}. Here, $\ket{\Psi[\bth]} = \hat{U}(\bth) \ket{\Psi_0}$ is obtained by application of a parametrized quantum circuit $\hat{U}(\bth)$ onto a reference state $\ket{\Psi_0}$. The cost function is evaluated on a quantum computer and the optimization is performed classically using $E(\bth)$ as input. The accuracy of VQE is therefore tied to the variational ansatz $\ket{\Psi[\bth]}$ and to the performance of the classical optimization, e.g., how often the cost function is called during the optimization and how well the approach converges to the global (as opposed to a local) minimum of $E(\bth)$.

\textbf{UCCSD Ansatz.}
The UCCSD ansatz takes the following form:
\bea
\ket{\Psi[\bth]} &=& e^{\hat{T}[\bth]-\hat{T}^\dagger[\bth]}\ket{\Psi_0} \notag \\ 
&=& e^{-i\sum_j \theta_j f_j(\{\hat{\sigma} \}) }\ket{\Psi_0}. \label{ucc}
\eea
The operator $\hat{T}[\bth]$ consists of single and double excitation operators with respect to the HF reference state $\ket{\Psi_0}$:
\be
\hat{T}[\bth] = \sum_{p\Bar{p}}\theta_{p}^{\Bar{p}}\cc_{\Bar{p}} \ca_{p} 
+ \sum_{p<q,\Bar{p}<\Bar{q}} \theta_{pq}^{\Bar{p}\Bar{q}}\cc_{\Bar{p}}\cc_{\Bar{q}}\ca_{q}\ca_{p}. 
\ee
Here $p, q$ and $\Bar{p}, \Bar{q}$ refer to the occupied and unoccupied MOs, respectively, with spin included implicitly. $f_j(\{\hat{\sigma} \})=\sum_{k}w_{j k} \hat{P}_k$ is a weighted sum of Pauli strings ($\hat{P}_k \in \{I, X, Y, Z \}^{\otimes N_q} $) for the qubit representation of the fermionic excitation operator associated with parameter $\theta_j$. Here $\theta_j$ run over the set of parameters $\theta_{p}^{\Bar{p}}$ and $\theta_{pq}^{\Bar{p}\Bar{q}}$. For the impurity model without spin-orbit interaction, only excitation operators which conserve respective number of electrons in the spin-up and spin-down sectors need to be considered. In practical implementation, a single step Trotter approximation is often adopted to construct the UCCSD circuit:
\be
\ket{\Psi[\bth]} \approx \prod_{jk} e^{-i \theta_j w_{jk}\hat{P}_{k}} \ket{\Psi_0}.
\ee
Furthermore, the final circuit state generally depends on the order of the unitary gates. In calculations reported here, we apply gates with single-excitation operators first following the implementation in Qiskit~\cite{Qiskit}. 

\textbf{Qubit-ADAPT VQE with simplified UCCSD pool.}
VQE-UCCSD is a useful reference point for quantum chemistry calculations. However, the fixed UCCSD ansatz has limited accuracy and often involves deep quantum circuits for implementations. Various approaches have been proposed to construct more compact variational ansatz with systematically improvable accuracy. In this work, we will focus on the qubit-ADAPT VQE method~\cite{MayhallQubitAVQE}, where the ansatz takes a similar pseudo-Trotter form:
\be
\ket{\Psi[\bth]} = \prod_{j=1}^{N_{\bth}} e^{-i \theta_j \hat{P}_{j}} \ket{\Psi_0}. \label{eq: qadapt_ansatz}
\ee
With qubit-ADAPT, the ansatz is recursively expanded by adding one unitary at a time, followed by reoptimization of parameters. The additional unitary is constructed with a generator selected from a predefined Pauli string pool which gives maximal energy gradient amplitude $|g|_\text{max}$ at the preceding ansatz state. The ansatz expansion process iterates until convergence, which is set by $|g|_\text{max} < 10^{-4}$ here. Note that we have set the half bandwidth of the original noninteracting lattice model to $D = 1$, such that $|g|_\text{max} \sim 0.1$~meV in physical systems with $D \sim$~1eV. 

The computational complexity of qubit-ADAPT VQE calculations is tied to the size of the operator pool, which consists of a set of Pauli strings. Naturally, one can construct an operator pool using all the Pauli strings in the qubit representation of fermionic single and double excitation operators. However, the dimension of this UCCSD-compatible pool is usually quite big and scales as $\bigO(N_q^4)$. Here we propose a much simplified operator pool, which consists of Pauli strings from single excitation and paired double excitation operators only. The pair excitation involves a pair of electrons with opposite spins, which are initially occupying the same spatial MO, hopping together to another initially unoccupied spatial MO. To further reduce the circuit depth, only one Pauli string is chosen from each qubit representation of the fermionic excitation operator. The qubit representation is a weighted sum of equal-length Pauli strings, and a specific choice of which one of them does not seem to be important in practical calculations reported here. This simplified pool containing operators arising from the UCC ansatz restricted to single and paired double excitation operators (sUCCSpD)\cite{oopUCCD,peUCC} greatly reduces the number of Pauli strings compared to the UCCSD pool. The dimension of this sUCCSpD pool scales as $\bigO(N_q^2)$. For the (2, 2) $e_g$ impurity model, the pool size reduces from 152 for UCCSD to 56 for sUCCSpD, and for the  (3, 3) $t_{2g}$ impurity model it reduces from 828 to 192. The code to perform the above qubit-ADAPT VQE calculations at statevector level with examples are available at figshare~\cite{pyqavqe}.

\textbf{Hamiltonian variational ansatz.}
The Hamiltonian sparsity in the AO basis naturally motivates the application of the Hamiltonian variational ansatz\cite{wecker2015_trotterizedsp}, which generally takes a form of multi-layer Trotterized annealing-like circuits. While different ways of designing specific HVA forms have been developed, we propose the following ansatz with $L$ layers for the impurity model:
\be
\ket{\Psi[\bth]} = \prod_{l=1}^{L} \prod_{j=1}^{N_\text{G}} e^{-i \theta_{lj} \hat{h}_{j}} \ket{\Psi_0}.
\ee
Here $\h = \sum_{j=1}^{N_\text{G}} \hat{h}_j$, with $\hat{h}_j$ being a subgroup of Hamiltonian terms which share the same coefficient and mutually commute. Such ansatz construction aims to differentiate the physical and bath orbitals, while retaining the degeneracy information among the orbitals in a systematic way. For each layer of unitaries, we first apply the multi-qubit rotations that are generated by the interacting part of the Hamiltonian, since these act as entangling gates. For the ($\M, \M$) impurity model, two reference states have been tried: $\ket{\Psi_0^\text{(I)}}$ is a simple tensor product state with $\M$ physical orbitals fully occupied and the bath orbitals empty; $\ket{\Psi_0^\text{(II)}}$ is the ground state of the noninteracting part of $\h$, which is equivalent to the one-electron core Hamiltonian in quantum chemistry. We did not find any significant difference between the two choices of reference state in practical simulations of the impurity models. Therefore, only HVA calculations with the reference state $\ket{\Psi_0^\text{(I)}}$ are reported here. We adopt the gradient-based Broyden–Fletcher–Goldfarb–Shanno (BFGS) algorithm as the classical optimizer. Proper parameter initialization for HVA optimization is crucial, as barren plateaus and local energy minima are generally present in the variational energy landscape. In practice, we find that a uniform initialization of the parameters, such as setting all to $\pi/7$, overall works well for simulations reported here. 

Inspired by the idea of adaptive ansatz generation~\cite{grimsleyAdaptiveVariationalAlgorithm2019}, we also tried constructing and optimizing an $L$-layer HVA ansatz by adaptively adding layers from $1$ to $L$. Specifically, the calculation starts with optimizing a single-layer ansatz, followed by appending another layer to the ansatz while keeping the first layer at previously obtained optimal angles. The two-layer ansatz is then optimized with the parameters for the new layer initialized randomly or uniformly. The procedure continues with the optimization of $l$-layer ansatz leveraging the $(l-1)$-layer solution until the ansatz reaches $L$ layers. 

Let the number of cost function evaluations for optimizing an $l$-layer ansatz be $N^{(2)}_l$. The total number of function evaluations amounts to $N^{(2)} = \sum_{l=1}^{L} N^{(2)}_l$. In practice, we find that the direct optimization of the $L$-layer ansatz using a uniform initialization takes $N^{(1)}$ function evaluations with $N^{(1)} \sim N^{(2)}_L < N^{(2)}$, and reaches the same accuracy. Starting with $L$ layers is therefore more efficient than growing the ansatz layer by layer. 

Intuitively, this can be related to the fact that successive HVA optimization introduce discontinuities in the variational path toward ground state whenever a new layer of unitaries is added. Since the energy gradient associated with new variational parameters that are initialized to zero (for continuity) vanishes (see Methods section), they have to be initialized away from zero. In other words, the $(l-1)$-layer HVA solution is not a good starting point for the optimization of the $l$-layer ansatz. The open source code to perform the above HVA calculations at the statevector level with examples are available at figshare~\cite{pyhva}.

\textbf{Hamiltonian commutator pool.\label{sec: HC}}
It has been demonstrated that the qubit-ADAPT VQE in the MO basis outperforms VQE-UCCSD calculations regarding circuit complexity and numerical accuracy~\cite{grimsleyAdaptiveVariationalAlgorithm2019, MayhallQubitAVQE}. Motivation by this observation, we compare the corresponding qubit-ADAPT VQE with Hamiltonian-compatible pool in AO basis and HVA calculations. Following HVA, we choose the simple tensor product state $\ket{\Psi_0^\text{(I)}}$ as the reference state. In qubit-ADAPT step, the energy gradient criterion $g_{\theta} = 2\Im[\Av{\Psi[\bth]}{\hat{P}\h}]$ to append a new unitary generated by $\hat{P}$ vanishes due to symmetry with $\Psi[\bth]$, if the number of Pauli-$Y$ operators in the Pauli string $\hat{P}$ is even~\cite{grimsleyAdaptiveVariationalAlgorithm2019, VQITE}. This can be simply shown from the following argument. Because the impurity model in this study respects time reversal symmetry and spin-flip ($Z_2$) symmetry, both Hamiltonian $\h$ and wavefunction are real ($\h=\h^{*}, \Psi[\bth]=\Psi[\bth]^{*}$). 
The Pauli string $\hat{P}$ is also real ($\hat{P}=\hat{P}^{*}$) if it has an even number of Pauli-Y operators.
Consequently, the expectation value of $\Av{\Psi[\bth]}{\hat{P}\h}$ is real and $g_{\theta}$ vanishes if the associated generator $\hat{P}$ has an even number of Pauli-$Y$ operators.


By construction, the sUCCSpD pool consists of Pauli strings of odd number of $Y$'s. However, the Hamiltonian of the impurity models studied here are all real. Consequently, all the Pauli strings in the qubit representation of the Hamiltonian contain an even number of $Y$'s, which excludes the option of directly constructing the operator pool from the Hamiltonian operators. Nevertheless, the practical usefulness of HVA implies that the Hamiltonian-like pool can be constructed by commuting the Hamiltonian terms, which we call Hamiltonian commutator (HC) pool $\mathscr{P}_\text{HC}$. Mathematically $\mathscr{P}_\text{HC}$ is constructed in the following manner,
\begin{align}
    \mathscr{P}_\text{HC}=\Big\lbrace \frac{1}{2i}[\hat{P},\hat{P}'] \,\,\big|\,\, \hat{P},\hat{P}'\in \mathscr{P}_\text{H}, \nonumber\\  \text{ and } N_{Y}([\hat{P},\hat{P}'])(\bmod\, 2) = 1\Big\rbrace,
\end{align}
Here $\mathscr{P}_{\text{H}}$ is the set of Pauli strings $\{\hat{P_h}\}$ present in the qubit representation of Hamiltonian  $\h=\sum_{h}w_{h}\hat{P}_{h}$. $N_{Y}(\hat{P})$ counts the number of $Y$ operators in the Pauli string $\hat{P}$.
Therefore, the size of $\mathscr{P}_\text{HC}$ can scale as $N_\text{H}^2$, where $N_\text{H}$ is the total number of Hamiltonian terms. Clearly, the pool $\mathscr{P}_\text{HC}$ should only be applied to sparse Hamiltonian systems. The dimension of the HC pool is 56 for the $e_g$ impurity model, and 192 for the $t_{2g}$ model.

\subsection{Quantum circuit implementation} \label{sec: implementation}
Performing a calculation on a quantum computer always needs to deal with the presence of noise. Even for ideal fault-tolerant quantum computers, quantum sampling (or shot) noise is present due to finite number of measurements that is used to estimate expectation values. The current noisy quantum devices exhibit additional noise originating from qubit relaxation and dephasing as well as hardware imperfections when implementing unitary gate operations. In this subsection, we describe several techniques adopted in our simulations to most efficiently use the available quantum resources and stabilize the calculations against sampling noise. We discuss how to mitigate gate noise in the final subsection.

\textbf{Measurement circuit reduction.}
The quantum circuit implementation for VQE and its adaptive version amounts to the direct measurement of the Hamiltonian as a weighted sum of Pauli string expectation values, $\av{\h}=\sum_{h}w_h \av{\hat{P}_h}$, with respect to parametrized circuits $U[\bth]$. Here, $\h=\sum_{h}w_h \hat{P}_h$ is the Hamiltonian in qubit representation. Because the number of shots (or repeated measurements) scales with the desired precision $\epsilon$ as $N_\text{sh} \propto \frac{1}{\epsilon^2}$ due to central limit theorem, $N_\text{sh}$ is often huge in practical calculations. Therefore, it is desirable to group the Pauli strings into mutually commuting sets such that the number of distinct measurement circuits is reduced to minimum. Indeed, many techniques to achieve such measurement reduction have been developed~\cite{MeasurementUnitaryPart2019, gokhale2019minimizing, zhao2020measurement,Crawford_2021, yen2021cartan, hamiltonianfactorization}. In this work, we adopt the measurement reduction strategy based on the Hamiltonian integral factorization~\cite{hamiltonianfactorization}, which shows a favorable linear system-size scaling of the number of distinct measurement circuits and embraces a diagonal representation for the operators to be measured. 

Specifically, we transform the physical subsystem Hamiltonian as follows:
\be
\h_\mathcal{S} = \sum_{\alpha\beta\s}\Tilde{\epsilon}_{\alpha\beta}\cc_{\alpha\s}\ca_{\beta\s} + \frac{1}{2}\sum_{\alpha\beta\gamma\delta}\sum_{\s\s'}V_{\alpha\beta\gamma\delta} \cc_{\alpha\s} \ca_{\beta\s} \cc_{\gamma\s'} \ca_{\delta\s'}, \label{eq: nn}
\ee
with $\Tilde{\epsilon}_{\alpha\beta}=\epsilon_{\alpha\beta}-\frac{1}{2}\sum_{\gamma}V_{\alpha\gamma\gamma\beta}$. A typical way to simplify the measurement of the two-body terms $\h_\mathcal{S}^{(2)}$ in Eq.~\eqref{eq: nn} is to perform nested matrix factorization for the Coulomb $V$ tensor. Namely, we first rewrite $\h_\mathcal{S}^{(2)}$ in the following factorized form by diagonalizing the real symmetric positive semidefinite supermatrix $V_{(\alpha\beta),(\gamma\delta)}$:
\be
\h_\mathcal{S}^{(2)} = \frac{1}{2}\sum_{l=1}^L\sum_{\alpha\beta}\sum_{\s} \left(\mathcal{L}_{\alpha\beta}^{(l)} \cc_{\alpha\s} \ca_{\beta\s}\right)^2. \label{eq: factor2}
\ee
Here $l$ runs through the $L$ positive eigenvalues of the supermatrix $V$, and the $l$th component of the auxiliary tensor $\mathcal{L}$ is obtained by multiplying the $l$th eigenvector with the square root of $l$th positive eigenvalue. Each tensor component, $\mathcal{L}^{(l)}$, which is a real symmetric matrix, is subsequently diagonalized to reach the following decomposition:
\bea
\sum_{\alpha\beta\s}\mathcal{L}^{(l)}_{\alpha\beta}\cc_{\alpha\s} \ca_{\beta\s} &=& \sum_{m=1}^{M_{l}}\lambda_m^{(l)} \sum_{\alpha\beta\s} U_{\alpha m}^{(l)}  U_{\beta m}^{(l)} \cc_{\alpha\s} \ca_{\beta\s} \notag \\
&=& \sum_{m=1}^{M_l}\sum_{\s}\lambda_m^{l} \hat{n}_{m\s}^{(l)} \label{eq: eigrep}
\eea
Here, we have defined $\hat{n}_{m\s}^{(l)} \equiv \sum_{\alpha\beta} U_{\alpha m}^{(l)}  U_{\beta m}^{(l)} \cc_{\alpha\s} \ca_{\beta\s}$.  The index $m$ goes through the $M_l$ nonzero eigenvalues $\lambda^{(l)}_m$ and associated eigenvectors $U_{m}^{(l)}$, which determines the single-particle basis transformation for the $l$th component. The whole embedding Hamiltonian of Eq.~\eqref{eq: h} can then be cast into the following doubly-factorized form with a unitary transformation similar to Eq.~\eqref{eq: eigrep} for the one-body part:
\bea
\h &=& \sum_{m=1}^{M_0}\sum_{\s}\epsilon^{(0)}_m\hat{n}_{m\s}^{(0)} +\frac{1}{2}\sum_{l=1}^{L}\sum_{m=1}^{M_l}\sum_{\s}\left(\lambda_{m}^{(l)}\hat{n}^{(l)}_{m\s}\right)^2, \label{eq: dfform} \nonumber \\
\eea
which is composed of $L+1$ groups characterized by unique single-particle basis transformations $\{U^{(l)} \}$, including one from the single-electron component. This form allows efficient measurement of the Hamiltonian expectation value using $L+1 \propto \bigO(N)$ distinct circuits for a generic quantum chemistry problem with single-particle basis dimension given by $N$.

The expectation value of $\h$ is obtained by measuring each group $l$ independently in the variational state $\ket{\Psi[\bth]}$. The variational state is transformed to the same representation used in the $l$th group by applying a series of Givens rotations, $\{e^{\theta_{\mu\nu} (\cc_{\mu\s}\ca_{\nu\s} -h.c.)} \}$, with the set of $\{\theta_{\mu\nu} \}$ determined by the single-particle transformation matrix $U^{(l)}$. Here $\mu$ and $\nu$ are generic indices for physical and bath orbital sites. Therefore, the number of distinct measurement circuits is $N_c = L+1$. As an example, we have $N_c=4$ for $e_g$ model. We refer to Methods section for further details.

In practice, it is advantageous to isolate the one-body and two-body terms that contain only density operators before the double factorization procedure, because they are already in a diagonal representation. For the $e_g$ model we have carried out the double-factorization with explicit calculations in Methods section and we ultimately find $N_{c}=3$ for the $e_g$ model. This can be compared with the Hamiltonian measurement procedure using the mutual qubit-wise commuting groups: operators that commute with respect to every qubit site are placed in the same group. This commuting Pauli approach generally needs $N_c \propto \bigO(N^4)$ distinct circuits for Hamiltonian measurement. And for the $e_g$ model, it requires $N_{c}=5$.

\textbf{Noise-resilient optimization.}
Although classical optimization approaches such as BFGS, which rely on a computation of the energy gradient, are effective, they rely on very accurate cost function evaluations. Because of the inherent noise in quantum computing, optimization algorithms that are robust to cost function noise are highly desirable. In the noisy quantum simulations reported here, we adopt two optimization techniques which are more tolerant to noise than BFGS: the sequential minimal optimization (SMO)\cite{nakanishi2020sequential} and \emph{Adadelta}~\cite{zeiler2012adadelta}. Because of their similar performance in the noisy simulations, we only discuss SMO in the main text, and leave the discussions of \emph{Adadelta} in Methods section.

SMO is the first technique we use for our noisy quantum simulations. Tailored to the qubit-ADAPT ansatz of Eq.~\eqref{eq: qadapt_ansatz} where each variational parameter is associated with a single Pauli string generator, the optimization consists of $N_\text{sw}$ sweeps of sequential single parameter minimization of the cost function. At a specific optimization step with varying parameter $\theta_j$ while keeping others fixed, the cost function has a simple form of $a\cos(2\theta_j - b) + c$, with the optimal $\theta_j^* = b/2$ if $a<0$ and $(b+\pi)/2$ otherwise. To determine the parameters $a,b,c$, one requires knowledge of function values for at least three mesh points in the range of $[-\pi/2, \pi/2)$. In practice, we use eight uniformly spaced mesh points to better mitigate the effect of noise in the cost function. Consequently, least square fitting is used to determine the values of $a, b$ and $c$. In SMO calculations, we use the number of sweeps as the parameter to control the convergence, which we set to $N_\text{sw} = 40$. Alternative control parameters, such as energy and gradient, usually are required to be evaluated at higher precision, which can be challenging and introduce additional quantum computation overhead. 

In this work, we perform noisy simulations with classical optimizations that include sampling noise due to finite number of measurements or shots ($N_\text{sh}$) 
as well as both sampling and gate noise. 
The purpose is to investigate the performance of the qubit-ADAPT algorithm in the presence of sampling and gate noise, and to separate the effects of sampling noise, which is controlled by a single parameter $N_{\text{sh}}$ from the effect of gate noise. The code with the circuit implementation of qubit-ADAPT VQE with examples on QASM simulator and quantum hardware are available at figshare~\cite{circqavqe}.

\begin{figure*}[t!]
	\centering
	\includegraphics[width=\linewidth]{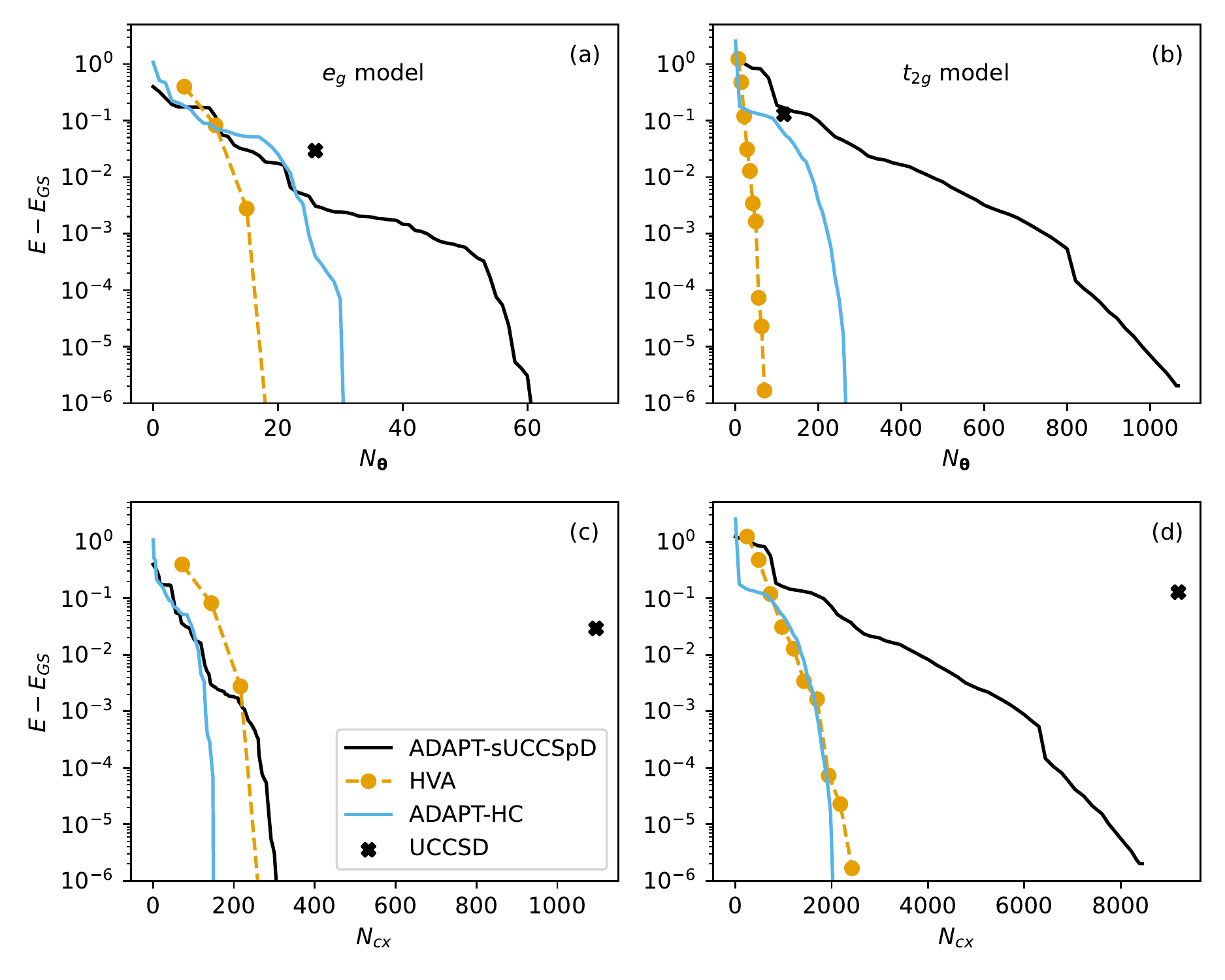}
	\caption{
	\textbf{Energy convergence of variational quantum eigensolver (VQE) calculations with four types of ans\"atze.} Panels (a,b) show the energy difference between the variational and the exact ground state energy $E_\text{GS}$ as a function of number of variational parameters $N_{\bth}$. Panels (c, d) show the energy difference versus the number of CNOT gates $N_\text{cx}$. Panels (a,c) are for the degenerate $(N_\mathcal{S}=2, N_\mathcal{B}=2)$ $e_g$ impurity model and panels (b,d) correspond to the (3, 3) $t_{2g}$ impurity model. VQE calculations are reported with fixed Hamiltonian variational ansatz (HVA, orange dashed line) and unitary coupled cluster ansatz with single and double excitations (UCCSD, black cross) as well as with adaptive ans\"atze constructed from a simplified unitary coupled cluster pool with single and paired double excitation operators (sUCCSpD, black line) and a Hamiltonian commutator pool (HC, sky blue line). Here $N_\text{cx}$ is estimated according to each multi-qubit rotation gate with a Pauli string generator $P$ of length $l$ contributing $2(l-1)$ CNOT gates, which assumes full qubit connection. The Hamiltonian parameters are $\epsilon=-9.8 (-12.7)$, $\lambda=0.3 (0.1)$, $\D=-0.3 (-0.3)$ with the same Hubbard $U=7$ for the $e_g$ ($t_{2g}$) model, corresponding to the correlated bad metallic regime. The energy unit is the half band width $D$ of the noninteracting DOS for the multi-band lattice model (see Fig.~\ref{fig: model}).
	}
	\label{fig: model23}
\end{figure*}

\begin{figure}[tbh]
	\centering
	\includegraphics[width=\linewidth]{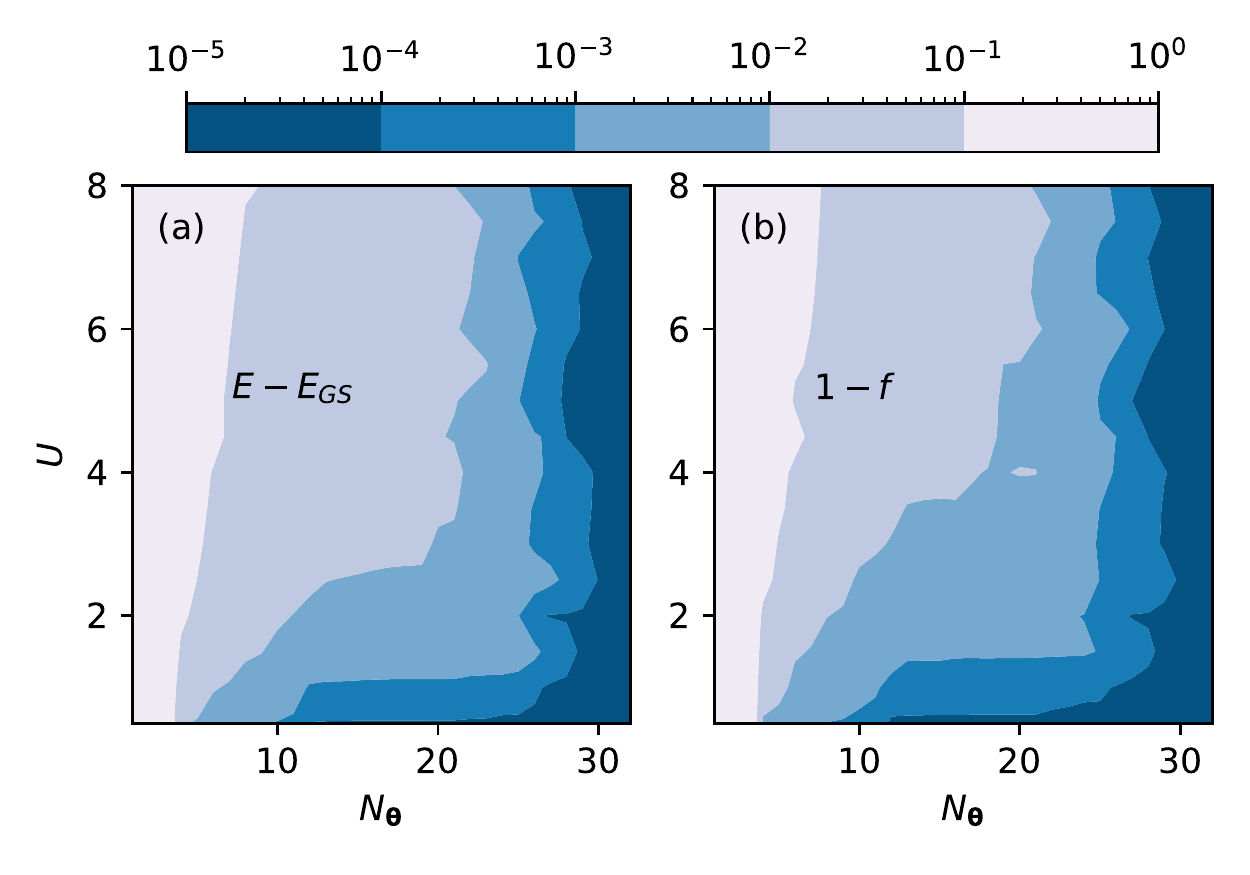}
	\caption{
	\textbf{Error and state fidelity analysis of qubit adaptive derivative-assembled pseudo-trotter (ADAPT) ansatz.} (a) Log-scale contour plot of the variational energy error $E-E_\text{GS}$ of the qubit-ADAPT ansatz as a function of $N_{\bth}$ and Hubbard $U$ for the (2, 2) $e_g$ impurity model. Here, $E_{\text{GS}}$ is the exact ground state energy and $E$ is the converged variational energy. (b) State infidelity $1-f=1-\abs{\ov{\Psi[\bth]}{\Psi_\text{GS}}}^2$ versus $N_{\bth}$ and $U$. Here, $\ket{\Psi[\bth]}$ is the converged ansatz state and $\ket{\Psi_{\text{GS}}}$ is the exact ground state. The color bar indicates a log scale from $10^{-5}$ to $1$. At a fixed energy accuracy, we find that $N_{\bth}$ generally increases with $U$ and then saturates. The same holds for the infidelity. We also observe a sharp rise of $N_{\bth}$ at smaller $U \approx 1 - 3$ when demanding an energy accuracy or infidelity below $10^{-3}$. This signifies the onset of correlation effects in the many-body ground state. The results are obtained from qubit-ADAPT calculations using Hamiltonian commutator pool of the $e_g$ model, where $U$ varies from $0.5$ to $8$ with $0.5$ as the step size. The other model parameters can be retrieved at figshare~\cite{pyqavqe}.
	}
	\label{fig: u}
\end{figure}

\subsection{Statevector simulations} \label{sec: sv}
In this section we present numerical simulation results using a statevector simulator, which is equivalent to a fault-tolerant quantum computer with an infinite number of measurements ($N_\text{sh}=\infty$). Figure~\ref{fig: model23} shows the ground state energy calculations of the (2, 2) $e_g$ and (3, 3) $t_{2g}$ impurity models using VQE-HVA as well as qubit-ADAPT VQE with sUCCSpD and HC pools. The reference UCCSD energy is $0.029$ higher than the exact ground state energy $E_\text{GS}$ for the $e_g$ model and $0.128$ higher for the $t_{2g}$ model. This implies that both models are in the strong electron correlation region. For calculations of the $e_g$ model, the energy converges below $10^{-5}$ with $N_{\bth} = 20$ variational parameters for VQE-HVA, $N_{\bth} = 59$ for ADAPT-sUCCSpD, and $N_{\bth} = 31$ for ADAPT-HC. Although the qubit-ADAPT VQE calculation on a statevector simulator is in principle deterministic, the operator selection from a predefined operator pool can introduce some randomness due to the numerical accuracy and near degeneracy of scores (i.e., the associated gradient components) for some operators. As a result, the converged $N_{\bth}$ can slightly change by about one between runs.

As a simple estimation of the circuit complexity for NISQ devices, we provide the number of CNOT gates $N_\text{cx}$ assuming full qubit connectivity, which can be realized in trapped ion systems. The converged circuit has $N_\text{cx}=288$ for VQE-HVA, $N_\text{cx}=292$ for ADAPT-sUCCSpD, and $N_\text{cx}=150$ for ADAPT-HC. As a reference, the UCCSD ansatz has $N_{\bth} = 26$ and $N_\text{cx}=1096$. The HVA calculation converges with the smallest number of variational parameters, but the number of CNOT gates ($N_\text{cx}$) is in between that of ADAPT-HC and ADAPT-sUCCSpD, because each variational parameter in HVA is associated with a generator composed of a weighted sum of Pauli strings. The ADAPT-HC calculation starts from a reference state $\ket{\Psi_0^\text{(I)}}$, a simple tensor product state in AO basis, with energy higher than the HF reference state used by ADAPT-sUCCSpD, yet ADAPT-HC converges faster to the ground state. In fact, the initial state fidelity, defined as $f \equiv \abs{\ov{\Psi_0}{\Psi_\text{GS}}}^2$, is 0.19 for ADAPT-HC, compared with 0.76 for ADAPT-sUCCSpD. Therefore, the final ansatz complexity does not show a simple positive correlation with the initial state fidelity, which implies that both the Hamiltonian structure and operator pool are determining factors.

Compared with ADAPT-sUCCSpD, the advantage of ADAPT-HC becomes more prominent when applied to the $t_{2g}$ model. To reach energy convergence below $10^{-5}$, ADAPT-HC needs $N_{\bth} = 270$ parameters and $N_\text{cx}=2052$ CNOTs, while ADAPT-sUCCSpD requires as many as $N_{\bth} = 1020$ parameters and $N_\text{cx}=8066$ CNOTs. For reference, the UCCSD ansatz has $N_{\bth} = 117$ parameters and $N_\text{cx}=9200$ CNOTs. The HVA calculation is carried out with up to $L = 10$ layers, which amounts to $N_{\bth} = 70$ and $N_\text{cx}=2420$, and the energy converges close to $10^{-6}$.

 We emphasize that strong electron correlation effects are present in our chosen model that lies deep in the bad metallic state~\cite{deMedici2011JanusfacedIO,Lanata-Hunds_metals-PRB-2013}. This state cannot be accurately captured within a mean-field description and hence requires the application of an appreciable number of unitary gates to the reference state. Generally, the circuit depth of a variational ansatz is tied to both the complexity of the problem (i.e. the complexity of the ground state wavefunction) and the desired state fidelity. As shown in Fig.~\ref{fig: u}, when we require a state fidelity close to $99.9\%$ or an energy error close to $0.001$, which is typically necessary in practical calculations, one observes a sharp rise of $N_{\bth}$ when the system is tuned from the weak correlation ($U<1$) to the strong correlation ($U>2$) regime by increasing Hubbard $U$. 

\subsection{Simulations with shot noise} \label{sec: qasm}
The ADAPT VQE calculations are often reported at the statevector level, and a systematic study including the effect of noise is not yet available~\cite{grimsleyAdaptiveVariationalAlgorithm2019, MayhallQubitAVQE, claudino2020benchmarking, yordanov2021qubit, bonet2021performance}. Here we present qubit-ADAPT VQE calculations of the $(2,2)$ $e_g$ model including shot noise.

\begin{figure*}[t]
	\centering
	\includegraphics[width=\linewidth]{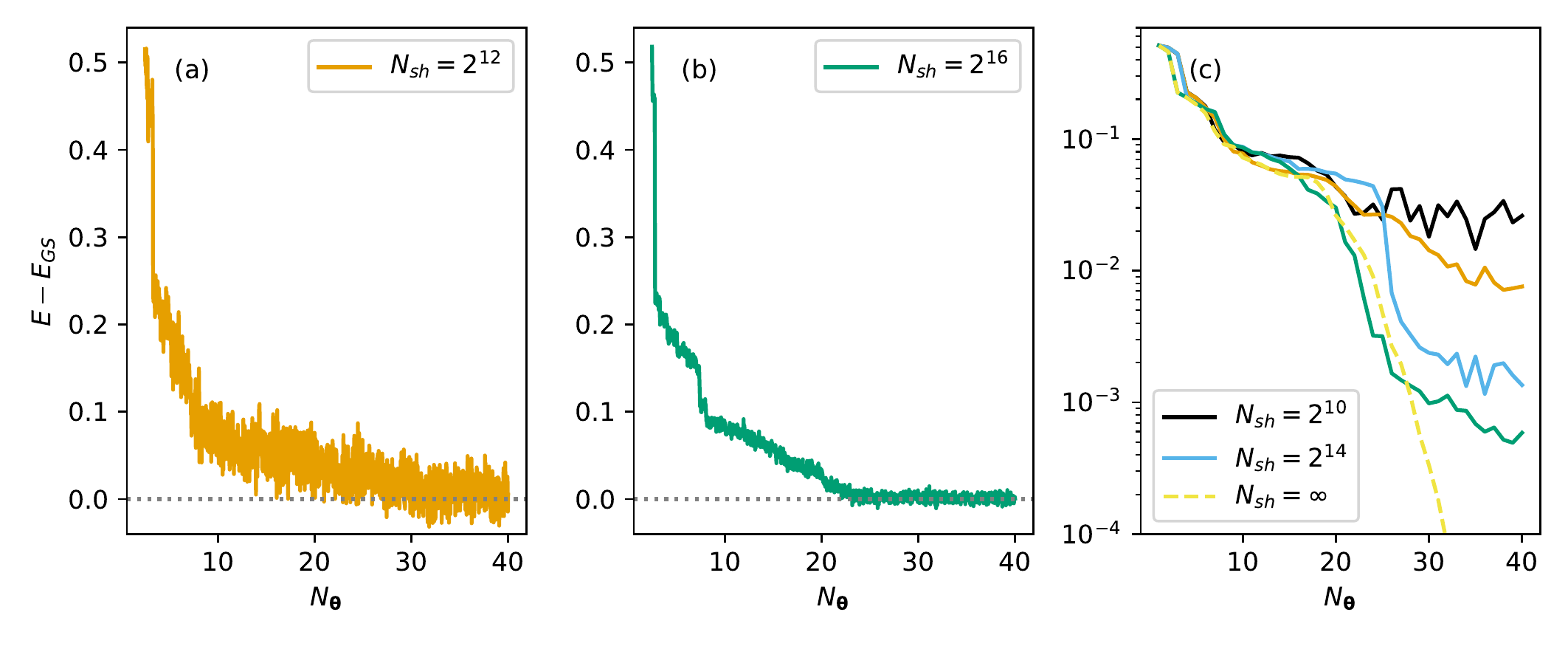}
	\caption{
	\textbf{Energy convergence of qubit adaptive derivative-assembled pseudo-trotter (ADAPT) simulations for $e_g$ model with shot noise.} The difference between the exact ground state energy $E_\text{GS}$ and that of qubit-ADAPT simulations with sampling noise, obtained with number of shots $N_\text{sh} = 2^{12}$ in panel (a) and $N_\text{sh} = 2^{16}$ in panel (b). Panel (c) shows the energy differences evaluated using statevector for the adaptive ans\"atze obtained in simulations including shot noise, with $N_\text{sh}=2^{10}$ (black line), $2^{12}$ (orange line), $2^{14}$ (sky blue line) and $2^{16}$ (bluish green line). The statevector simulation results ($N_\text{sh}=\infty$, yellow line) of the qubit-ADAPT algorithm are also shown in dashed line for reference. Hamiltonian parameters are identical to those used in Fig.~\ref{fig: model23}.
	}
	\label{fig: smo}
\end{figure*}

Figure~\ref{fig: smo} shows the representative convergence behavior of the qubit-ADAPT energy with an increasing number of variational parameters $N_{\bth}$ calculated using different number of shots per observable measurement: panel (a) is for $N_\text{sh} = 2^{12}$, and panel (b) is for $N_\text{sh} = 2^{16}$. We use SMO for the classical optimization. The adaptive ansatz energy $E$ overall decreases as the circuit grows and more variational parameters are used. The energy uncertainty is tied to the number of shots $N_\text{sh}$. The energy spread roughly reduces by a factor of 4 when $N_\text{sh}$ increases from $2^{12}$ (a) to $2^{16}$ (b), consistent with the 16-fold increase in $N_\text{sh}$ due to central limit theorem. 

The energy points shown include not only the final SMO optimized energies of the qubit-ADAPT ansatz with $N_{\bth}$ parameters, but also the intermediate energies after each of the $N_\text{sw}=40$ sweeps during SMO optimizations to provide more detailed convergence information. The above reported $N_\text{sh}$ is referred to measurements for SMO optimizations. At the operator screening step of the qubit-ADAPT calculation to expand the ansatz by appending an additional optimal unitary, we fix $N_\text{sh}=2^{16}$ shots for energy evaluations in all cases, and determine the energy gradient by the parameter-shift rule~\cite{mari2021estimating}. 

To further assess the quality of the qubit-ADAPT ansatz obtained in these QASM simulations, we plot in Fig.~\ref{fig: smo}(c) the ansatz energies evaluated using a statevector simulator at the end of each noisy SMO optimization. The four solid curves are calculated using the variational parameters that are obtained by QASM optimizations with different numbers of shots $N_\text{sh}$ as indicated and noiseless optimization results are shown for comparison as the dashed line. While there is no clear order of the energies during early stages of the simulation, the final convergence is consistently improved with more shots. Specifically, the error converges close to and below $10^{-3}$ for $N_\text{sh}=2^{14}$ and $2^{16}$ and the fidelity $f$ improves beyond $99.9\%$. The associated single-particle density matrix elements also converge to an accuracy better than $10^{-2}$.

Similar QASM simulations of qubit-ADAPT VQE have been performed using the \emph{Adadelta} optimizer, as specified in Methods section. Generally we find the numerical results and the dependence on the number of shots to be comparable to SMO. Compared with SMO, \emph{Adadelta} can potentially take advantage of multiple QPUs by evaluating the gradient vector in parallel.

\begin{figure*}[t]
	\centering
	\includegraphics[width=\linewidth]{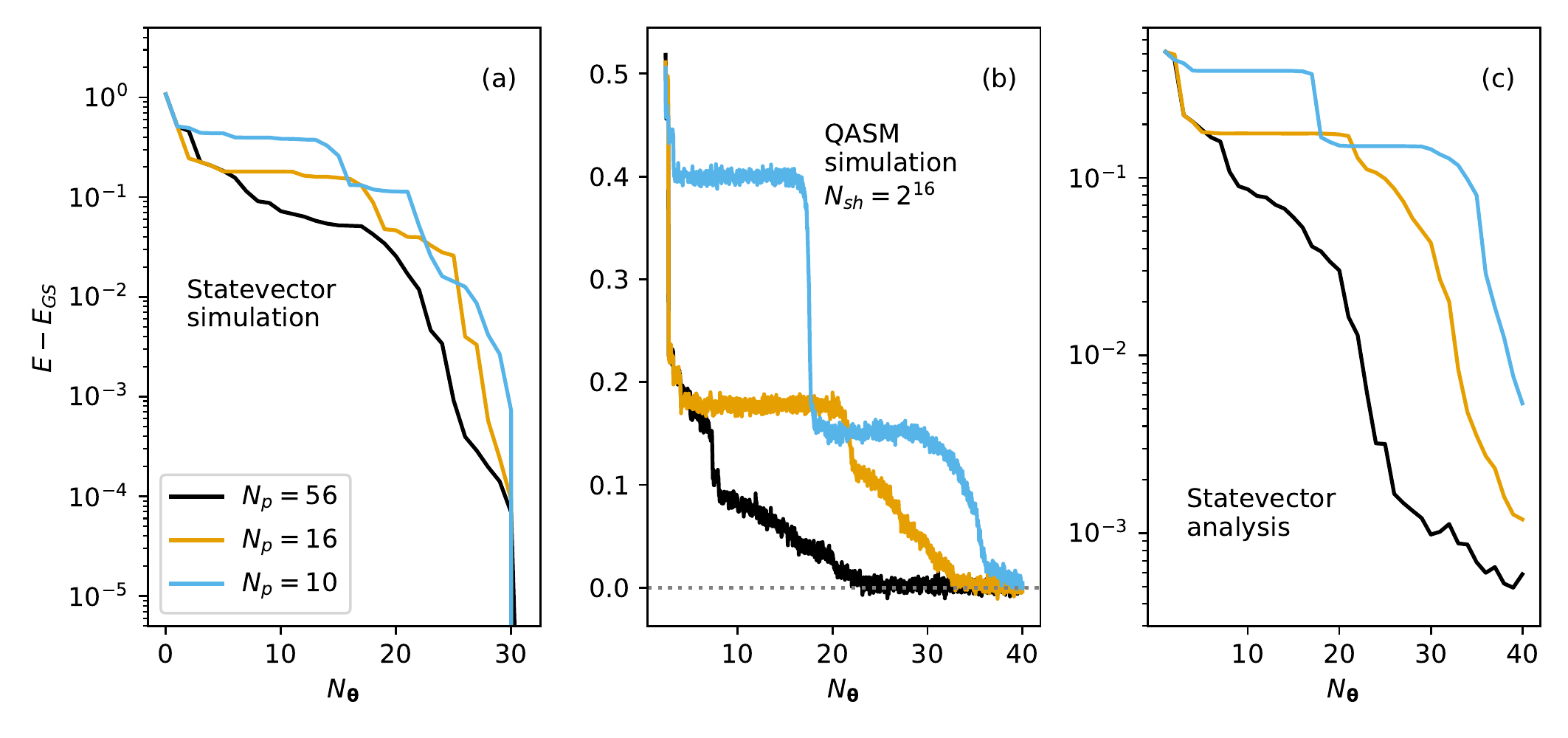}
	\caption{
	\textbf{Pool size dependence of the energy convergence behavior for qubit adaptive derivative-assembled pseudo-trotter (ADAPT) calculations of the $e_g$ model.} The difference between the exact ground state energy $E_\text{GS}$ and qubit-ADAPT results as a function of $N_{\bth}$ from (a) statevector simulations and (b) Quantum assembly language (QASM)-based simulations with $N_\text{sh}=2^{16}$ shots using three different operator pools of size 56 (black line), 16 (orange line) and 10 (sky blue line), derived from the Hamiltonian commutator pool. The respective energy differences evaluated using statevector for the adaptive ans\"atze obtained in the noisy simulations of panel (b) are shown in panel (c).
	}
	\label{fig: pools}
\end{figure*} 

\subsection{Discussion of optimal pool size} \label{sec: discussion}
One important factor determining the computational load of qubit-ADAPT VQE calculations is the size of the operator pool $N_\text{p}$. One simple strategy to reduce $N_\text{p}$ is to strip off Pauli $Z$'s in the pool of operators, because they contribute negligibly to the ground state energy as pointed out in Refs.~\cite{MayhallQubitAVQE, yordanov2021qubit}. This reduces $N_\text{p}$ of the Hamiltonian commutator (HC) pool from 56 to 16 for the $e_g$ model, and from 192 to 60 for the $t_{2g}$ model, due to a large degeneracy. Furthermore, some qualitative guidance has been laid out in the literature to construct a minimal complete pool (MCP) of size $2(N_q-1)$~\cite{MayhallQubitAVQE, shkolnikov2021avoiding}, where $N_q$ is the number of qubits. Indeed, we find that a MCP can be constructed using a subset of operators in HC pool. 

We discover a dichotomy that the reduction of the pool size can potentially make the optimization of the qubit-ADAPT ansatz more challenging, especially in the presence of noise. Figure~\ref{fig: pools} compares qubit-ADAPT calculations using three different pool sizes of dimension 56, 16 and 10, which were introduced above. Figure~\ref{fig: pools}(a) shows the qubit-ADAPT energies with increasing $N_{\bth}$ from statevector simulations of the $e_g$ model using the three pools.  All the simulations converge with 31 parameters and final CNOT gate numbers $N_\text{cx} = 150,\, 98, $ and $62$ that decrease for the smaller pools. The details of convergence rate of the three runs differ significantly. When the pool dimension decreases, the region of $N_{\bth}$ with minimal energy change expands, as seen by the almost flat segments of the curves of Fig.~\ref{fig: pools}(a). The minimal energy gain implies that small noise in the cost function evaluation could deteriorate the parameter optimization. 

Indeed as shown in Fig.~\ref{fig: pools}(b), the qubit-ADAPT energy from noisy simulation converges slower as the pool size decreases. The flat segments in the energy curves become more evident owing to the stochastic energy errors. We further analyse the quality of the qubit-ADAPT ansatz by evaluating the energy at optimal angles obtained in noisy simulations, as plotted in Fig.~\ref{fig: pools}(c). The energy difference is 0.001, 0.027, 0.135 at $N_{\bth}=31$ where the statevector simulation converges, and 0.0006, 0.001, 0.005 at the end of $N_{\bth}=40$ for calculations with pools of size 56, 16 and 10, respectively. 

Our analysis clearly shows the strikingly distinct convergence behaviors of qubit-ADAPT calculations using different complete operator pools in the presence of sampling noise. This indicates that the optimal pool in practical calculations can be a trade-off between choosing a small pool size and guaranteeing sufficient connectivity of the operators in the pool.

\begin{figure}
    \centering
    \includegraphics[width=\columnwidth]{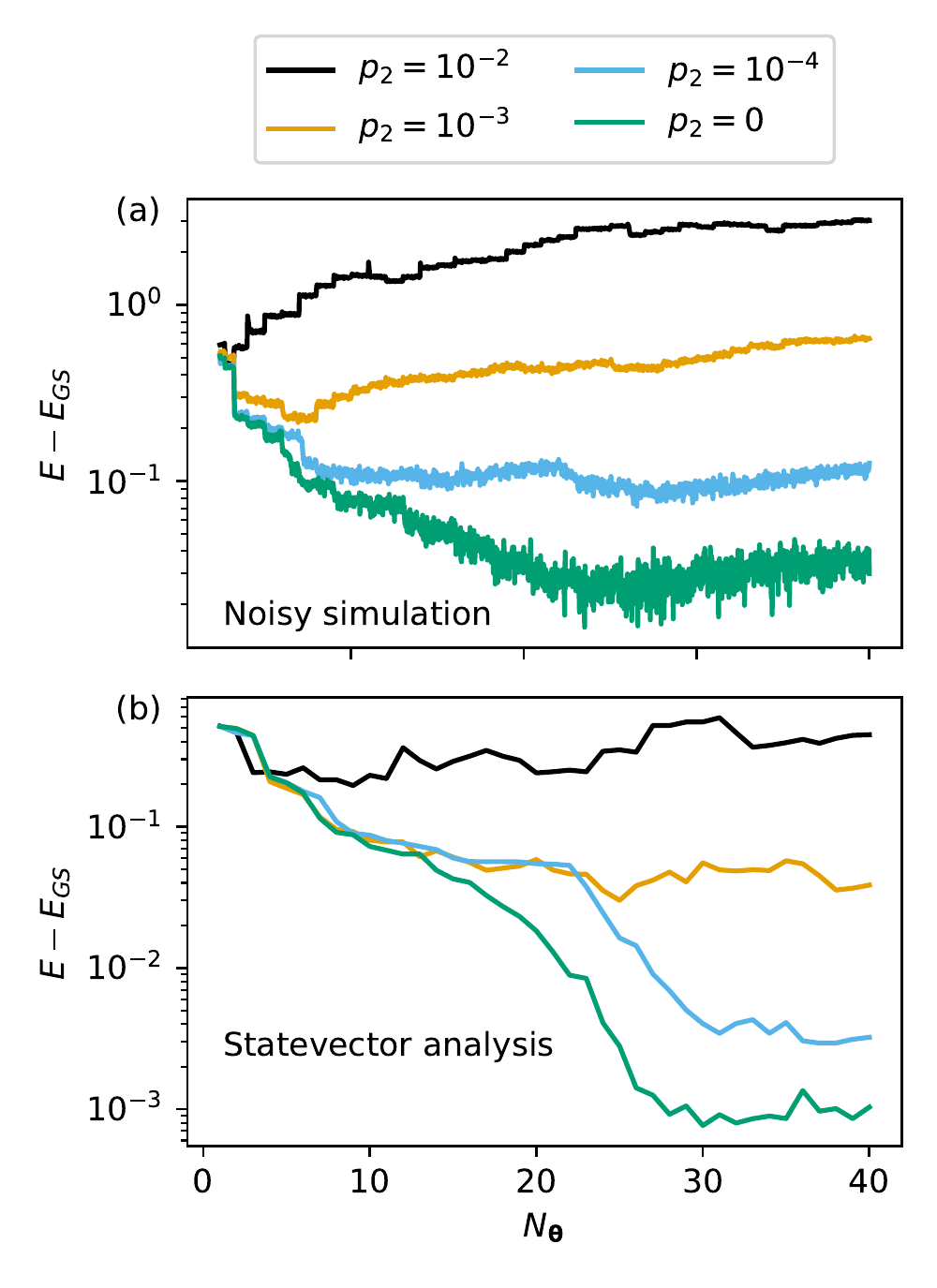}
    \caption{
    \textbf{Noisy qubit adaptive derivative-assembled pseudo-trotter (ADAPT) simulations and analysis of $e_g$ model.} (a) Difference between the exact ground state energy $E_\text{GS}$ and qubit-ADAPT noisy simulation results with a uniform two-qubit gate error rate $p_2=10^{-1}$ (black line), $10^{-3}$ (orange line), $10^{-4}$ (sky blue line) and $0$ (bluish green line). We use a uniform single-qubit error rate $p_1 = 10^{-4}$ and $N_{\text{sh}} = 2^{16}$ shots per measurement circuit. (b) Energy differences evaluated using statevector for the adaptive ans\"atze obtained in the noisy simulations. The noisy simulations are performed with the Hamiltonian commutator pool of size $56$.
    } 
    \label{fig: nm}
\end{figure}

\subsection{Simulations with noise models \label{sec:noise_models}}
Besides the inherent sampling noise in quantum computing, NISQ hardware is subject to various other error effects. These include coherent errors due to imperfect gate operations as well as stochastic errors due to qubit decoherence, dephasing and relaxation. Here, we perform a preliminary investigation of the impact of hardware imperfections on qubit-ADAPT VQE calculations by adopting a realistic decoherence noise model proposed by Kandala et al. in Ref.~\cite{hardware_efficient_vqe}. The model includes an amplitude damping channel ($\rho \to \sum_{i=1}^2 E_i^a \rho E_i^{a\dagger}$) and a dephasing channel ($\rho \to \sum_{i=1}^2 E_i^d \rho E_i^{d\dagger}$). These act on the qubit density matrix following each single-qubit or two-qubit gate operation. The Kraus operators are given as:
\bea
E_1^a &=& 
\begin{pmatrix}
1 & 0\\
0 & \sqrt{1-p^a}
\end{pmatrix}, 
E_2^a = 
\begin{pmatrix}
0 & \sqrt{p^a}\\
0 & 0
\end{pmatrix}, \nonumber \\
E_1^d &=& 
\begin{pmatrix}
1 & 0\\
0 & \sqrt{1-p^d}
\end{pmatrix}, 
E_2^d = 
\begin{pmatrix}
0 & 0\\
0 & \sqrt{p^d}
\end{pmatrix}.
\eea
The error rates $p^a = 1-e^{-\tau/T_1}$ and $p^d = 1-e^{-2\tau/T_{\phi}}$ are determined by the gate time $\tau$, the qubit relaxation time $T_1$ and the dephasing time $T_{\phi}=2T_1T_2/(2T_1-T_2)$, where $T_2$ is the qubit coherence time. For the sake of simplicity of the analysis, we choose a uniform single-qubit gate error rate $p_1^a = p_1^d \equiv p_1 = 10^{-4}$, which is close to the value found in current hardware. We also assume a uniform two-qubit error rate  $p_2^a = p_2^d = p_2$ that we vary between $10^{-4}$ and $10^{-2}$, in order to study the impact of two-qubit noise on the VQE optimization.

Figure~\ref{fig: nm}(a) shows a typical qubit-ADAPT energy curve $E-E_\text{GS}$ during optimization as a function of the number of variational parameters $N_{\bth}$ obtained in noisy simulations with $p_2 = 10^{-2}$, $10^{-3}$, and $10^{-4}$. Here, $E_\text{GS}$ is the exact ground state energy. The results with only single-qubit noise are also shown for reference. Figure~\ref{fig: nm}(b) contains the associated exact energies for the ansatz states, which we obtain by evaluating the VQE ansatz on a statevector simulator.

For $p_2 = 10^{-2}$, which represents the current hardware noise level, the noisy energy in panel (a) increases with $N_{\bth}$, indicating that the error rate is too large to get reliable energy estimation. Nevertheless, as shown in the corresponding statevector analysis in panel (b), one still observes a sizable energy reduction in the early stage of the optimization, where the ansatz state fidelity improves from $0.19$ in the initial state to $\approx 0.70$ when $4 < N_{\bth} < 9$ (not shown). When further increasing $N_{\bth}$, however, the statevector ansatz energy shows an upward trend due to noise accumulation, signifying a failure of the noisy optimization. For a smaller error rate $p_2 = 10^{-3}$, which was demonstrated recently with the IBM Falcon device~\cite{cnot999}, the noisy energy in panel (a) initially decreases and reaches a minimum near $N_{\bth}=7$. This is again followed by an upturn as the number of variational parameters $N_{\bth}$ grows. On the other hand, the corresponding statevector analysis in panel (b) shows a clear continuous energy improvement up to $N_{\bth}=25$, followed by a saturation with small fluctuations. The ansatz state fidelity saturates near $0.97$ (not shown). Similar observations apply to the noisy simulations with other two-qubit error rates. The statevector analysis shows that the energy converges at an error $\approx 3\times 10^{-3}$ with a fidelity $\approx 0.997$ for $p_2 = 10^{-4}$. When including only single-qubit errors, we find an error $\approx 1\times 10^{-3}$ with a fidelity $\approx 0.9992$. 

The observed improvement of the ansatz (revealed using statevector analysis), even though the noisy energy expectation value increases, is intriguing. This effect is most clearly seen in results for $p=10^{-3}$ between $7 \leq N_{\theta} < 25$. It demonstrates a robustness of VQE to certain types of noise effects and can be rationalized as follows. Assuming for simplicity a global depolarizing error channel, we can relate the expectation value of an observable $\bar{\av{O}}$ with respect to a noisy density matrix to the noiseless result $\av{O}$ as $\bar{\av{O}} = (1-p)\av{O} + \frac{p}{2^n}\Tr[O]$~\cite{Urbnek2021MitigatingDN, Vovrosh2021SimpleMO}. Since any observable can be shifted to be traceless ($\Tr[O]=0$), $\av{O}$ is equivalent to $\bar{\av{O}}$ up to a constant scaling factor. The noise thus only rescales the energy landscape of the variational ansatz, and maintains the optimal parameters. The fact that we find the ansatz energy to saturate in the statevector analysis with finite $p_2$ is caused by our choice of noise model, which includes noise effects beyond a global depolarizing channel. This observation of state improvement during optimization masked by noisy energy expectation values suggests that with reasonably small error rates, expensive error mitigation techniques may be restricted to the final converged state at the end of VQE calculations to ensure accurate observable measurements.

\begin{figure*}
    \centering
    \includegraphics[width=\textwidth]{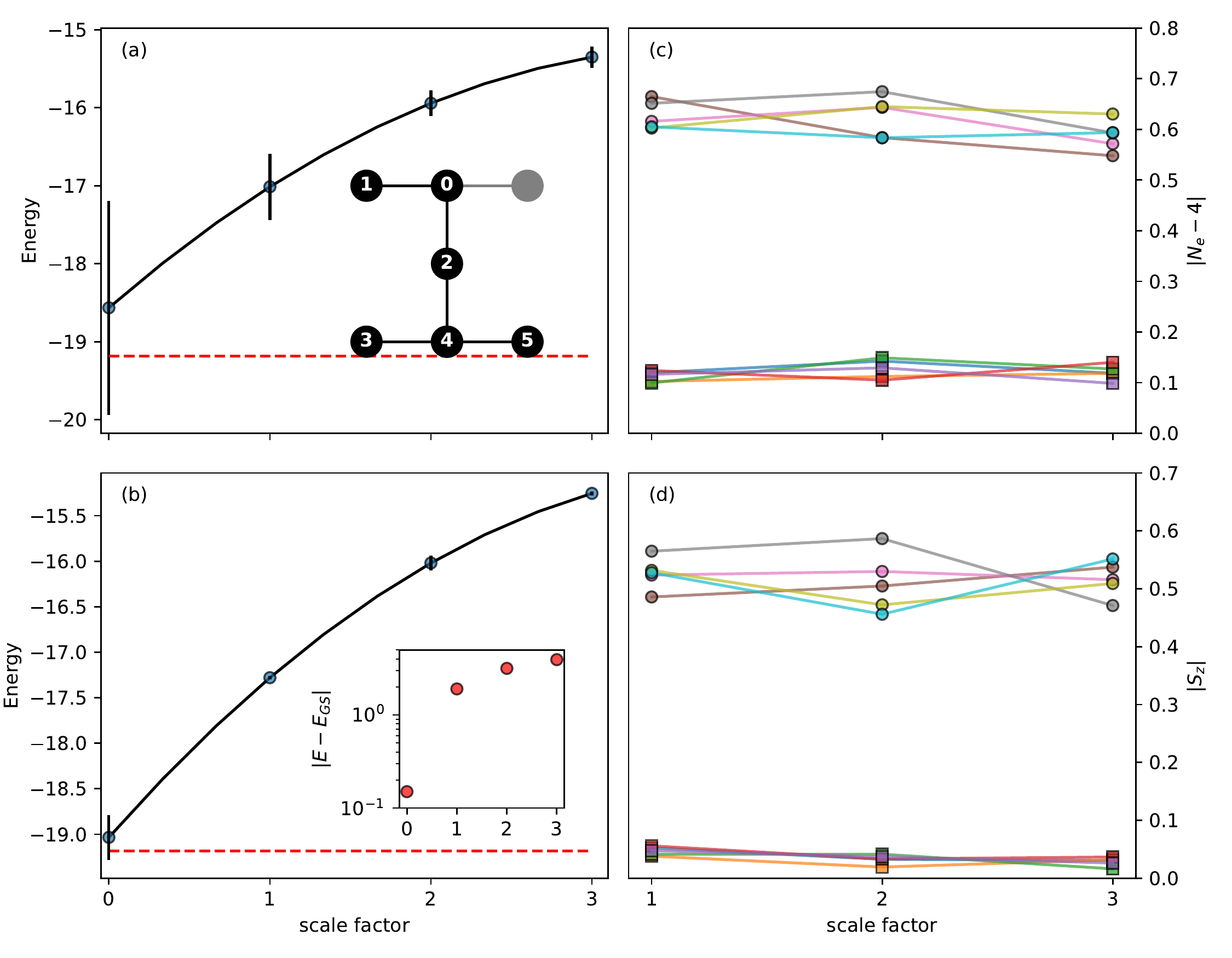}
    \caption{
    \textbf{Estimating ground state energy of the $e_g$ model on IBM device \texttt{ibmq\_casablanca}.} Richardson energy extrapolation is applied by a quadratic curve fitting for three data points of increasing noise scale with averages over 10 runs in (a) and an optimal subset of 5 runs in (b). Distinct but equivalent hardware native circuits are associated with each run owing to the nondeterministic nature of local random unitary folding and transpilation. The average number of electrons $N_e$ and total spin z-component $S_z$ for each of the 10 runs in terms of their deviations from ideal values are plotted in (c) and (d), respectively. The optimal subset of 5 runs are identified by smaller symmetry violations $\abs{N_e - 4}<0.2$ and $\abs{S_z} < 0.1$. The inset in (a) shows the qubit layout of \texttt{ibmq\_casablanca}. The dark numbered circles represent the qubits adopted in the calculation with that particular order. Inset in (b): the energy error $\Delta(E)=\abs{E-E_\text{GS}}$ in log scale. The error bar denotes the standard deviation of the sample mean.
    } 
    \label{energy_estimation}
\end{figure*}

\subsection{Estimating ground state energy on NISQ devices} \label{sec: device}
As a further step to benchmark the realistic noise effect on qubit-ADAPT VQE calculations of the multi-orbital quantum impurity models, we measure the Hamiltonian expectation value of the $e_g$ model with a converged qubit-ADAPT ansatz on the IBM quantum device \texttt{ibmq\_casablanca}. The ansatz with optimal parameters is obtained with the HC pool using statevector simulations. The converged qubit-ADAPT ansatz used for the ground state energy estimate has 32 parameters, and the associated 32 generators for multi-qubit unitary gates are listed in Methods section.

To reduce the noise in the cost function measurement, it is essential to utilize a range of error mitigation techniques. We employ the standard readout error mitigation using the full confusion matrix approach, as implemented in Qiskit~\cite{Qiskit}. The adopted measurement circuits based on Hamiltonian integral factorization also allows convenient symmetry detection and filtering with respect to how well the ansatz preserves the total electron number $N_e=4$ and total spin z-projection $S_z=0$. The gate error is mitigated using zero noise extrapolation (ZNE) with Richardson second-order polynomial inference~\cite{VDynamics_Li, ZNE_Temme}. The noise scale factor increases from 1 to 2 and 3 for each measurement circuit by local random unitary folding following the implementation in Mitiq~\cite{larose2020mitiq, ZNE_digital}. Because of the random gate folding and the stochastic SWAP mapping during transpilation to native gates~\cite{Qiskit}, we perform ten runs for each measurement circuit at each noise level to smooth out the nondeterministic effects with averaging. For each run, we apply $N_{\text{sh}} = 2^{14}$ shots for the measurements.

Figure~\ref{energy_estimation}(a) shows the Richardson extrapolation for the ground state energy with measured points at noise scale factors $\lambda = 1, 2, 3$, taking all ten runs for each $\lambda$ into account. The estimated energy has an absolute error $\Delta(E) = 0.6\pm 1.4$ compared with the exact result indicated by the horizontal dashed line. This corresponds to a relative error of $3\%$. The standard deviation is obtained by fitting the sample points with a second order polynomial using the SciPy function \texttt{curve\_fit} which takes both the mean values and standard deviations into account~\cite{2020SciPy-NMeth}. In the postprocessing for the mean value of the energy cost function from statistical samplings, we first apply readout calibration, followed by symmetry filtering which discards the configurations with total electron number $N_e \neq 4$ or total spin $S_z \neq 0$. We observe that the ten runs can be divided into two groups based on the average $N_e$ and $S_z$ evaluated before symmetry filtering, as shown in Fig.~\ref{energy_estimation}(c) and (d). A subgroup of five runs denoted by square symbols have much less bias away from the correct conserved quantum numbers $N_e = 4$ and $S_z = 0$ than the other five runs shown as circles. A more accurate ground state energy can be obtained when restricting to this optimal subgroup, as shown in Fig.~\ref{energy_estimation}(b). The estimated energy error reduces significantly to $\Delta(E) = 0.1\pm 0.2$, with a relative error of $0.7\%$.

In the above calculations on QPU, the circuits are transpiled into the basis gates of \texttt{ibmq\_casablanca} device using the qubit layout and coupling map illustrated in the inset of Fig.~\ref{energy_estimation}(a). Due to the limited qubit connectivity between nearest neighbours, each of the three transpiled measurement circuits for the $e_g$ model contains about 350 CNOT gates, which amounts to over two-fold increase compared to about 150 CNOTs without qubit swapping. Therefore, we also benchmark the calculations on other types of QPUs with full qubit connectivity such as trapped-ion devices. As an initial reference, we perform an energy estimation with the same ansatz on Quantinuum's trapped-ion Honeywell System Model H1-2. The transpiled circuits have about 150 two-qubit ZZMax gates as expected. Due to limited access to the device, we apply only $N_\text{sh}=450$ shots per circuit for the measurements without utilizing any error mitigation. The energy thus obtained is $-17.6\pm 2$, which should be compared with data points in Fig.~\ref{energy_estimation}(a) at a scale factor 1, and is found to be located near the lower end of that range. Here the error bar is estimated using multiple runs of simulations with the associated system Model H1-2 emulator (H1-2E) including a realistic noise model.

\section{Conclusions} \label{sec: conclusion}
In an effort towards performing hybrid quantum-classical simulations of realistic correlated materials using a quantum embedding approach~\cite{dmft_georges96, dmft_kotliar06, sun2016quantum, knizia2012dmet, ga_uo2, ga_dmet, gqce, sakurai2021hybrid, Vorwerk2022QuantumET}, we assess the gate depth and accuracy of variational ground state preparation with fixed and adaptive ans\"atze for two representative interacting multi-orbital, $e_g$ and $t_{2g}$, impurity models. To take advantage of the sparsity of the Hamiltonian in the atomic orbital representation in real space, we consider the HVA ansatz and an adaptive variant in the qubit-encoded atomic orbital basis. A HC pool composed of pairwise commutators of the Hamiltonian terms is developed to allow fair comparison between the qubit-ADAPT and HVA ansatz. For reference, the standard UCCSD and related qubit-ADAPT calculations using UCCSD-compatible pools are also presented. The qubit-ADAPT calculation with a HC pool generally produces the most compact circuit representation with a minimal number of CNOTs in the final converged circuit. The fixed HVA ansatz follows very closely and has the additional advantage of requiring the least variational parameters $N_{\bth}$. 

To address the effect of quantum shot noise, we report QASM simulations of qubit-ADAPT VQE in the presence of shot noise for different numbers of shots ($N_\text{sh}$) that allows controlling the stochastic error. For our benchmark, we adopt state-of-the-art techniques such as low-rank tensor factorization to reduce the number of distinct measurement circuits and a noise resilient optimization including sequential minimal optimization and \emph{Adadelta}. We find a modest number of shots $N_\text{sh}=2^{14}$ per measurement circuit can lead to a variational representation of the ground state with fidelity $f>99.9\%$.

We further discuss ways to simplify the pool operators and reduce the pool size using $e_g$ model as an example. It is pointed out that a minimal complete pool, as defined in Ref.~\cite{MayhallQubitAVQE, shkolnikov2021avoiding}, can be constructed using a subset of the HC pool. While a simplified pool can reduce the quantum computation resource in the adaptive operator screening procedure, it can make the classical optimization more complicated, especially in the presence of noise. This suggests both the dimension and connectivity of operators are joint determining factors to design a practically optimal pool.

To assess the effects of realistic noise on VQE calculations of multi-orbital impurity models, we perform qubit-ADAPT VQE calculations with a realistic decoherence noise model that includes amplitude and dephasing error channels. We find the impact of two-qubit errors to dominate over those of single-qubit errors, also since they are larger in NISQ hardware. We report that practically useful results can be obtained for $p_2 = 10^{-3}$, which is close to current hardware levels. Importantly, we observe that the classical optimization continues to improve the ansatz even in a regime, where the noisy energy expectation value starts to rise. We reveal this behavior by executing the ansatz state on statevector simulators. Such persisting ansatz state improvement masked by noise shows that VQE is robust to certain noise effects and implies that costly error mitigation methods can potentially be reserved for the evaluation of expectation values in the final converged state. 

Finally, we measure the energy for a converged qubit-ADAPT ansatz of the $e_g$ model on the \texttt{ibmq\_casablanca} QPU and Quantinuum's H1-2 device. Using the results from IBM hardware, we obtain an error of $0.1$ (0.7\%) for the total energy by adopting error mitigation techniques such as zero-noise extrapolation, combined with a careful post-selection based on symmetry and the conservation of quantum numbers.

Moving forward, the full qubit-ADAPT VQE calculations of quantum impurity models will be extended from noisy QASM simulations to simulations that include device specific noise effects beyond our decoherence model and finally to experiments on real hardware. Our study shows that an array of error mitigation techniques, including readout calibration, zero-noise extrapolation~\cite{VDynamics_Li, ZNE_Temme}, and potentially probabilistic error cancellation~\cite{ZNE_Temme, endo2018practical, Berg2022ProbabilisticEC}, Clifford data regression~\cite{cdr, lowe2021unified}, and probabilistic machine learning based techniques~\cite{ rogers2021error}, need to be adopted to reach sufficiently accurate results. This is especially important when using VQE as an impurity solver in a quantum embedding approach as sufficiently accurate impurity model results are needed in order to enable convergence of the classical self-consistency loop. Our results constitute an important step forward in demonstrating high fidelity ground state preparation of impurity models on quantum devices. This is essential for realizing correlated material simulations through hybrid quantum-classical embedding approaches, where the ground state preparation of a generic $f$-electron impurity model consisting of 28 spin-orbital is at the verge of achieving practical quantum advantage~\cite{gqce}.

\section{Methods}
\subsection{Energy gradient of HVA \label{app: hva}}
Here we show that the outermost $l$th layer gradient component vanishes ($\left. \frac{\partial \mathcal{E}(\bth)}{\partial \theta_{l j}}\right|_{\bth_{l} = 0}=0$) for an $l$-layer HVA ansatz $\ket{\Psi_l[\bth]} = \Pi_{j=1}^{N_\text{G}} e^{-i\theta_{l j} \hat{h}_j} \ket{\Psi_{l-1}[\bth]}$: 
\bea
\left. \frac{\partial \mathcal{E}(\bth)}{\partial \theta_{l j}}\right|_{\bth_{l} = 0} &=& \left. \frac{\partial \Av{\Psi_{l}[\bth]}{\h } }{\partial \theta_{l j}}\right|_{\bth_{l} = 0} \nonumber \\
&=& -i \Av{\Psi_{l-1}[\bth]}{\h \hat{h}_j } + c.c. \,
\eea
Because the system Hamiltonian $\h$ under study is real due to time-reversal symmetry, HVA is also real by construction. Therefore, $\Av{\Psi_{l-1}[\bth]}{\h \hat{h}_j }$ is real, and $\left. \frac{\partial \mathcal{E}(\bth)}{\partial \theta_{l j}}\right|_{\bth_{l} = 0}$ vanishes. Note that the exactly same reason motivates the development of HC pool for qubit-ADAPT calculations.

\subsection{Hamiltonian factorization of the impurity model}
\label{double-factorization-e_g-model}
Here we explain explicitly how the Hamiltonian factorization is obtained using the $e_{g}$ model as an example, whose Hamiltonian takes the following specific form:
\bea
    \h &=& D\sum_{i=1}^{2}\sum_{\s}\left(\cc_{i\s}\fa_{i\sigma}+h.c.\right) \label{eq: hybridization} \\ 
    &+&J/2\left(\cc_{1\up}\ca_{2\up} + \cc_{1\dw}\ca_{2\dw} + h.c.\right)^2 \label{eq: pair} \\ 
    &+&U\sum_{i=1}^{2}\hat{n}_{i\up}\hat{n}_{i\dw} +(U-2J)\sum_{\s\s'}\hat{n}_{1\s}\hat{n}_{2\s'} \\
    &+&\Tilde{\epsilon} \sum_{i=1}^2\sum_{\s}\hat{n}_{i\s} + \lambda\sum_{i=1}^2\sum_{\s}\hat{n}^{f}_{i\s}.
\eea
Here $\hat{n}_{i\s}=\cc_{i\s}\ca_{i\s}$ and $\hat{n}_{i\s}^f=\fc_{i\s}\fa_{i\s}$ are the electron occupation number operators for the physical and bath orbitals, respectively. The factorization procedure is only needed for the single-particle hybridization term~\eqref{eq: hybridization} and the pair hopping and spin flip terms~\eqref{eq: pair}, as the rest are already in the diagonal representation.

The hybridization term~\eqref{eq: hybridization} can be written in a diagonal form through single-particle rotations on the physical and bath orbitals as follows:
\be
    \sum_{i=1}^2\left(\cc_{i\s}\fa_{i\s}+h.c.\right)=-\hat{n}^{(0)}_{1\s} -\hat{n}^{(0)}_{2\s} +\hat{n}^{(0)}_{3\s} +\hat{n}^{(0)}_{4\s} ,\label{hybridization terms}
\ee
where $\hat{n}^{(0)}_{m\s}=\hat{c}^{\dagger(0)}_{m\s}\hat{c}^{(0)}_{m\s}$ and the rotated fermionic operators $\hat{c}^{(0)}_{m\s}$ are given by,
\bea
\hat{c}^{(0)}_{1\s}&=&\frac{1}{\sqrt{2}}(\hat{c}_{1\s}+\hat{f}_{1\s}), \hat{c}^{(0)}_{2\s}=\frac{1}{\sqrt{2}}(\hat{c}_{2\s}+\hat{f}_{2\s}), \nonumber \\
\hat{c}^{(0)}_{3\s}&=&\frac{1}{\sqrt{2}}(\hat{c}_{1\s}-\hat{f}_{1\s}), \hat{c}^{(0)}_{4\s}=\frac{1}{\sqrt{2}}(\hat{c}_{2\s}-\hat{f}_{2\s}). \label{eq: basis0}
\eea
This can be derived conveniently in the matrix formulation:
\bea
&&\sum_{i=1}^2\left(\cc_{i\s}\fa_{i\s}+h.c.\right) \nonumber \\
&&= 
\begin{pmatrix}
\cc_{1\s} & \cc_{2\s} & \fc_{1\s} & \fc_{2\s}
\end{pmatrix}
\begin{pmatrix}
0 & 0 & 1 & 0 \\
0 & 0 & 0 & 1 \\
1 & 0 & 0 & 0 \\
0 & 1 & 0 & 0
\end{pmatrix}
\begin{pmatrix}
\ca_{1\s} \\
\ca_{2\s} \\
\fa_{1\s} \\
\fa_{2\s}
\end{pmatrix}
\nonumber \\
&&= 
\begin{pmatrix}
\cc_{1\s} & \cc_{2\s} & \fc_{1\s} & \fc_{2\s}
\end{pmatrix}
\begin{pmatrix}
\frac{1}{\sqrt{2}} & 0 & \frac{1}{\sqrt{2}} & 0 \\
0 & \frac{1}{\sqrt{2}} & 0 & \frac{1}{\sqrt{2}} \\
\frac{1}{\sqrt{2}} & 0 & -\frac{1}{\sqrt{2}} & 0 \\
0 & \frac{1}{\sqrt{2}} & 0 & -\frac{1}{\sqrt{2}}
\end{pmatrix}
\nonumber \\
&&\times
\begin{pmatrix}
-1 & 0 & 0 & 0 \\
0 & -1 & 0 & 0 \\
0 & 0 & 1 & 0 \\
0 & 0 & 0 & 1
\end{pmatrix}
\begin{pmatrix}
\frac{1}{\sqrt{2}} & 0 & \frac{1}{\sqrt{2}} & 0 \\
0 & \frac{1}{\sqrt{2}} & 0 & \frac{1}{\sqrt{2}} \\
\frac{1}{\sqrt{2}} & 0 & -\frac{1}{\sqrt{2}} & 0 \\
0 & \frac{1}{\sqrt{2}} & 0 & -\frac{1}{\sqrt{2}}
\end{pmatrix}
\begin{pmatrix}
\ca_{1\s} \\
\ca_{2\s} \\
\fa_{1\s} \\
\fa_{2\s}
\end{pmatrix}
\nonumber \\
&&=
\begin{pmatrix}
\hat{c}^{\dagger(0)}_{1\s} & \hat{c}^{\dagger(0)}_{2\s} & \hat{c}^{\dagger(0)}_{3\s} & \hat{c}^{\dagger(0)}_{4\s}
\end{pmatrix}
\begin{pmatrix}
-1 & 0 & 0 & 0 \\
0 & -1 & 0 & 0\\
0 & 0 & 1 & 0 \\
0 & 0 & 0 & 1
\end{pmatrix}
\begin{pmatrix}
\hat{c}^{(0)}_{1\s} \\
\hat{c}^{(0)}_{2\s} \\
\hat{c}^{(0)}_{3\s} \\
\hat{c}^{(0)}_{4\s}
\end{pmatrix}. \nonumber \\ \label{eq: facrtor1}
\eea
The pair hopping and spin flip terms of the second line of Eq.~\eqref{eq: pair} can be rewritten as:
\be
J/2\left(
\begin{pmatrix}
\cc_{1\up} & \cc_{1\dw} & \cc_{2\up} & \cc_{2\dw}
\end{pmatrix}
\mathcal{L}^{(1)}
\begin{pmatrix}
\ca_{1\up} \\
\ca_{1\dw} \\
\ca_{2\up} \\
\ca_{2\dw}
\end{pmatrix}
\right)^2.
\ee
with 
\be
\mathcal{L}^{(1)} =
\begin{pmatrix}
0 & 0 & 1 & 0 \\
0 & 0 & 0 & 1 \\
1 & 0 & 0 & 0 \\
0 & 1 & 0 & 0
\end{pmatrix}.
\ee
The above expression is obtained by diagonalizing the Coulomb supermatrix of $V_{(\alpha\beta), (\gamma\delta)}$ with density-density elements set to zero, $V_{(\alpha\alpha), (\gamma\gamma)}\equiv 0$, which gives a single eigenvector associated with nonzero eigenvalue. Following the similar derivation in Eq.~\eqref{eq: facrtor1}, the pair hooping and spin flip terms have the following diagonal representation:
\be
J/2\left(-\hat{n}_{1\up}^{(1)} -\hat{n}_{1\dw}^{(1)} + \hat{n}_{2\up}^{(1)} +\hat{n}_{2\dw}^{(1)}\right)^2,
\ee
with $\hat{n}^{(1)}_{m\s}=\hat{c}^{\dagger(1)}_{m\s} \hat{c}^{(1)}_{m\s}$ and 
\be
\hat{c}_{1\s}^{(1)}=\frac{1}{\sqrt{2}}(\ca_{1\s}+\ca_{2\s}), \hat{c}_{2\s}^{(1)}=\frac{1}{\sqrt{2}}(\ca_{1\s}-\ca_{2\s}). \label{eq: basis1}
\ee

Finally, we can represent the embedding Hamiltonian for $e_{g}$ model in the following doubly-factorized form:
\bea
    \h &=& D\sum_{\s}\left( -\hat{n}^{(0)}_{1\s} -\hat{n}^{(0)}_{2\s} +\hat{n}^{(0)}_{3\s} +\hat{n}^{(0)}_{4\s} \right) \nonumber \\ 
    &+&J/2\left( -\hat{n}_{1\up}^{(1)} -\hat{n}_{1\dw}^{(1)} + \hat{n}_{2\up}^{(1)} +\hat{n}_{2\dw}^{(1)} \right)^2 \nonumber \\ 
    &+&U\sum_{i=1}^{2}\hat{n}_{i\up}\hat{n}_{i\dw} +(U-2J)\sum_{\s\s'}\hat{n}_{1\s}\hat{n}_{2\s'} \nonumber \\
    &+&\Tilde{\epsilon} \sum_{i=1}^2\sum_{\s}\hat{n}_{i\s} + \lambda\sum_{i=1}^2\sum_{\s}\hat{n}^{f}_{i\s}.
\eea

With the Hamiltonian integral factorization we find that three distinct measurement circuits are needed for the Hamiltonian expectation value: (i) the diagonal terms in the original atomic orbital basis, (ii) the hybridization terms in the basis of $c^{(0)}_{m\s}$~\eqref{eq: basis0}, (iii) the pair hopping and spin flip terms in the basis of $c^{(1)}_{m\s}$~\eqref{eq: basis1}.

\subsection{Quantum Simulation with Adadelta optimizer}\label{sec: adadelta}
\begin{figure*}[t!]
	\centering
	\includegraphics[width=\linewidth]{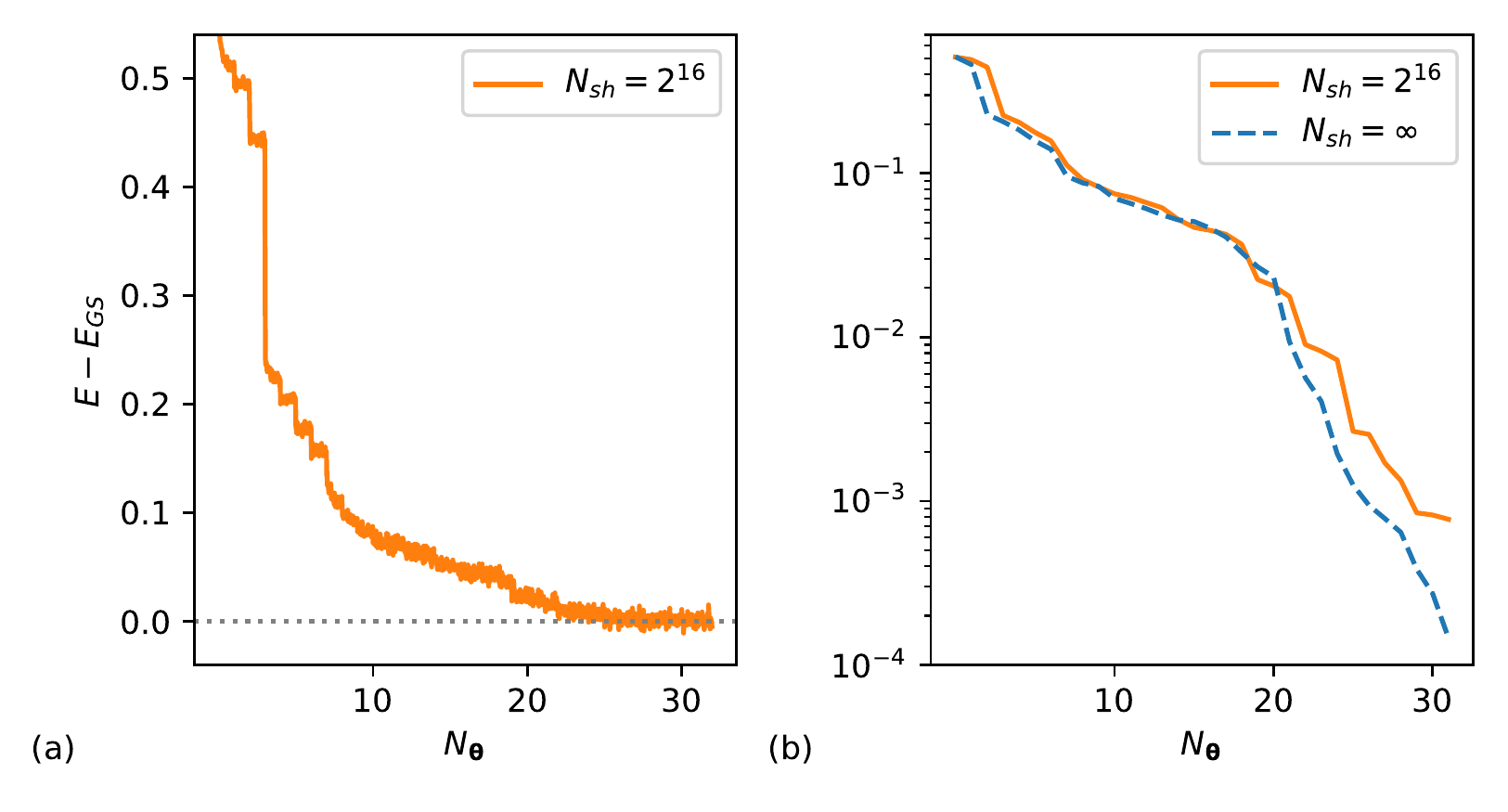}
	\caption{
	\textbf{Energy convergence of qubit adaptive derivative-assembled pseudo-trotter (ADAPT) noisy simulations of $e_g$ model with \emph{Adadelta} optimizer.} The difference between the exact ground state energy $E_\text{GS}$ and qubit-ADAPT noisy simulation results, obtained with number of shots $N_\text{sh} = 2^{16}$ in panel (a). The energy differences evaluated using statevector for the adaptive ans\"atze obtained in the noisy simulations with $N_\text{sh}=2^{16}$ (orange line) are shown in panel (b). The statevector simulation results ($N_\text{sh}=\infty$) of the qubit-ADAPT method are also shown in blue dashed line for reference.
	}
	\label{fig: adadelta}
\end{figure*}
In the main text, we reported the qubit-ADAPT VQE calculation with shots using the SMO optimizer. Here we additionally perform  the calculations using the $\emph{Adadelta}$ optimization method, which is potentially tolerant to cost function errors~\cite{zeiler2012adadelta}. Below we describe the implementation of the algorithm followed by the results. 

The algorithm minimizes the cost function along the steepest decent direction in parameter space, with a parameter update at step $t$ as $\bth_t = \bth_{t-1} - \mathbf{w}_t\odot \mathbf{g}_t$. The gradient vector is determined from the derivative of the energy function along every parameter direction $\mathbf{g}_{t}=\nabla_{\bth}E(\bth_{t})$, where $E(\bth)=\Av{\Psi[\bth]}{\h}$ is the estimated energy. The set of parameter-dependent adaptive learning rates are determined as $\mathbf{w}_t = \frac{\sqrt{\Delta \bth_{t-1} + \epsilon}}{\sqrt{\mathbf{s}_t + \epsilon}}$, where the leaked average of the square of rescaled gradients at the previous step is obtained as $\Delta \bth_{t-1} = \beta \Delta \bth_{t-2} + (1-\beta)(\mathbf{w}_{t-1}\odot \mathbf{g}_{t-1})^2$, and that of gradients is evaluated as $\mathbf{s}_{t} = \beta \mathbf{s}_{t-1} + (1-\beta)\mathbf{g}_t^2$. The operator $\odot$ denotes element-wise product. The \emph{Adadelta} algorithm involves a hyperparameter $\epsilon$ to regularize the ratio in determining $\mathbf{w}_t$, which is set to $10^{-8}$, and a mixing parameter set to $\beta = 0.9$. The leaked averages are all initialized to zero. We fix the number of steps in \emph{Adadelta} optimization to $N_{s}=250$ in our simulations. Considering that the evaluation of one gradient component associated with a variational parameter involves cost function measurements at two distinct parameter points following the parameter-shift rule, the quantum computational resource for \emph{Adadelta} optimization is comparable to SMO with $N_\text{sw}=60$.

Figure~\ref{fig: adadelta} shows the representative convergence behavior of qubit-ADAPT energy with increasing number of variational parameters $N_{\bth}$ calculated using number of shots $N_\text{sh} = 2^{16}$ per observable. The adaptive ansatz energy $E$ decreases as the circuit depth increases with more variational parameters. The energy points shown include not only final \emph{Adadelta} optimized energies of the qubit-ADAPT ansatz with $N_{\bth}$ parameters, but also intermediate energies for the $250$ Adadelta steps to provide a detailed view of the convergence. For the operator screening step of the qubit-ADAPT calculation we fix $N_\text{sh}=2^{16}$ for energy evaluations in all cases, and determine the energy gradient by the parameter-shift rule~\cite{mari2021estimating}. The final energy error from the calculations with \emph{Adadelta} is $E-E_{GS}=4.4\times 10^{-3}$. This is comparable with the result from SMO optimizer.

\subsection{The ground state ansatz of (2, 2) $e_g$~model used on \texttt{ibmq\_casablanca}}
\label{sec: operators_ansatz}
The qubit-ADAPT ansatz takes the pseudo-Trotter form. The converged ansatz for the $e_g$ model which we used for the calculations on IBM quantum hardware \texttt{ibmq\_casablanca} is composed of 32 generators for the multi-qubit unitary gates, which are listed here with parity encoding (in the order that they appear in the ansatz):
\begin{itemize}
    \item[] IIIZXY, IYXIII, XYZIII, IIZYXZ, IXYIII,
    \item[] ZXYIIZ, XYIIZZ, XYIIIZ, IIIIYX, IZXYXX,
    \item[] IIXZYI, IIXIIY, IIXIZY, IIZYXZ, IIIZYX,
    \item[] YXIIII, IZXIZY, IIXIZY, IIYIIX, IIZXYI,
    \item[] IZXYXX, IZYIZX, ZYXIII, ZYIIZX, IIYIIX,
    \item[] IIIIXY, IIXIIY, IIXIYZ, IIXZYI, YXXIZX,
    \item[] IIXIYZ, YXXIIX.
\end{itemize}

\section*{Data availability}
All the data to generate the figures are available at figshare~\cite{dataegt2gmoidel}. Data supporting the calculations are available together with the codes at figshare~\cite{pyhva, pyqavqe, circqavqe}. All other data are available from the corresponding authors on reasonable request.

\section*{Code availability}
All the computer codes developed and used in this work are available open-source at figshare~\cite{pygqce, pyhva, pyqavqe, circqavqe}.

\bibliography{refabbrev, ref}

\section*{Acknowledgements}
The authors acknowledge valuable discussions with Thomas Iadecola, Niladri Gomes, Cai-Zhuang Wang and Nicola Lanat\`a. 
This work was supported by the U.S. Department of Energy (DOE), Office of Science, Basic Energy Sciences, Materials Science and Engineering Division, including the grant of computer time at the National Energy Research Scientific Computing Center (NERSC) in Berkeley, California. The research was performed at the Ames Laboratory, which is operated for the U.S. DOE by Iowa State University under Contract No. DE-AC02-07CH11358. 
We acknowledge use of the IBM Quantum Experience, through the IBM Quantum Researchers Program. The views expressed are those of the authors, and do not reflect the official policy or position of IBM or the IBM Quantum team. This research also used resources of the Oak Ridge Leadership Computing Facility, which is a DOE Office of Science User Facility supported under Contract DE-AC05-00OR22725.

\section*{Author contributions}
A.M. and Y.Y. developed the codes and performed the simulations. N.F.B. and J.C.G. contributed to the HVA calculations. A.M., P.P.O., and Y.Y. analyzed the results. Y.Y., A.M. and P.P.O. wrote the paper with input from all  authors. Y.Y. and P.P.O. conceived and supervised the project.

\section*{Competing interests}
The authors declare no competing interests.

\end{document}